\documentclass[a4paper,11pt]{article}
\newcommand{\ctns}{\mathcal{K}}
\newcommand{\cder}{\mathcal{D}}

\pdfoutput=1 

\usepackage{jheppub} 
\newcommand{\la}{\langle}
\newcommand{\ra}{\rangle}
\usepackage[T1]{fontenc} 
\usepackage{tikz}
\allowdisplaybreaks

\title{\boldmath Multi-spin soft bootstrap and scalar-vector Galileon}

\author[a]{Karol Kampf,}
\author[a]{Jiří Novotný,}
\author[a]{Filip Přeučil,}
\author[b,a]{Jaroslav Trnka}

\affiliation[a]{Institute of Particle and Nuclear Physics, Charles University,\\ V Holešovičkách 2, 180 00 Prague 8, Czech Republic}
\affiliation[b]{Center for Quantum Mathematics and Physics (QMAP),\\Department of Physics, University of California, Davis, CA 95616, USA}

\emailAdd{karol.kampf@mff.cuni.cz}
\emailAdd{jiri.novotny@mff.cuni.cz}
\emailAdd{filip.preucil@mff.cuni.cz}
\emailAdd{trnka@ucdavis.edu}

\abstract{We use the amplitude soft bootstrap method to explore the space of effective field theories (EFT) of massless vectors and scalars. It is known that demanding vanishing soft limits fixes uniquely a special class of EFTs: non-linear sigma model, scalar Galileon and Born-Infeld theories. Based on the amplitudes analysis, we conjecture no-go theorems for higher-derivative vector theories and theories with coupled vectors and scalars. We then allow for more general soft theorems where the non-trivial part of the soft limit of the $(n{+}1)$-pt amplitude is equal to a linear combination of $n$-pt amplitudes. We derive the form of these soft theorems for general power-counting and spins of particles and use it as an input into the soft bootstrap method in the case of Galileon power-counting and coupled scalar-vector theories. We show that this unifies the description of existing Galileon theories and leads us to the discovery of a new exceptional theory: Special scalar-vector Galileon.} 

\begin{document} 
\maketitle
\flushbottom

\section{Introduction}
\label{sec:intro}

On-shell scattering amplitudes are fundamental objects in weakly coupled quantum field theories. The textbook approach to calculate amplitudes uses the method of Feynman diagrams which is a diagrammatic representation of the perturbative expansion of the path integral. In the last few decades, we have seen many discoveries of hidden mathematical structures and symmetries in the perturbative S-matrix which are completely invisible in the standard Feynman diagrams approach. Unitarity methods \cite{Bern:1994cg,Bern:1994zx} and recursion relations exploit the locality and unitarity of tree-level amplitudes \cite{Britto:2004ap,Britto:2005fq,Cohen:2010mi,Cheung:2015cba,Cheung:2015ota} and loop integrands \cite{ArkaniHamed:2010kv,Baadsgaard:2015twa}, CHY formula evaluates tree-level amplitudes as worldsheet integrals \cite{Cachazo:2013hca,Cachazo:2013iea}, color-kinematics duality relates amplitudes in different theories simplifying significantly complex calculations \cite{Bern:2008qj,Bern:2010ue,Bern:2019prr}, and in certain theories the S-matrix has been formulated in geometric language using positive geometries leading to Amplituhedron picture for planar ${\cal N}=4$ SYM amplitudes \cite{Arkani-Hamed:2016byb,Arkani-Hamed:2013kca,Arkani-Hamed:2013jha,Arkani-Hamed:2014dca,Arkani-Hamed:2017vfh,Arkani-Hamed:2018rsk,Damgaard:2019ztj,Herrmann:2020qlt}, and beyond \cite{Arkani-Hamed:2017mur,Bern:2015ple,Arkani-Hamed:2019mrd,Trnka:2020dxl,Jagadale:2021iab}.

Surprisingly, some of these recent developments have also been found in the context of tree-level amplitudes in effective field theories. In the extension of tree-level unitarity and BCFW recursion relations, the scattering amplitudes in non-linear sigma model, Dirac-Born-Infeld theory, Volkov-Akulov, and Galileons were uniquely specified by special soft limit behavior \cite{Kampf:2012fn,Kampf:2013vha,Bijnens:2019eze,Cheung:2014dqa,Cheung:2016drk,Elvang:2018dco} and further lead to the discovery of Special Galileon theory \cite{Cheung:2014dqa,Cachazo:2014xea}. In all cases, the theory is uniquely fixed by vanishing soft limit of tree-level scattering amplitudes. Interestingly, same theories also appear in the context of color-kinematics duality, CHY construction and ambitwistor strings \cite{Low:2019wuv,Low:2020ubn,Cachazo:2014xea,Casali:2015vta}, and belong to a larger web of dualities among massless tree-level amplitudes \cite{Cheung:2017ems,Bern:2019prr}. In the more generalized setup, we can replace the requirement of vanishing soft limits by non-trivial (but known) soft theorems. Such soft theorems appeared in the past in the context of supergravity, amplitudes on the Coulomb branch, or in the context of low energy limit of string theory \cite{Huang:2015sla,DiVecchia:2015jaq,Bianchi:2016viy,Green:2019rhz,Wang:2015aua}, and more recently in the $SU(N)/SU(N-1)$ non-linear sigma model \cite{Kampf:2019mcd}. Similar strategy has also been used in fixing amplitudes using IR constraints, gauge invariance, large-$z$ behavior or imposing BCJ relations \cite{Arkani-Hamed:2016rak,Rodina:2016mbk,Rodina:2016jyz,Rodina:2018pcb,Carrasco:2019qwr}. The amplitudes approach was also useful in the classification of counterterms and higher derivative operators in non-linear sigma model, gauge theory and gravity \cite{Carrasco:2019yyn,Dai:2020cpk,Rodina:2021isd,Carrasco:2021ptp}. 

In this paper, we continue the analysis of EFTs using amplitude methods. We first explore the space of all vector effective field theories with vanishing soft limits. At the leading derivative order this produces Born-Infeld as a unique theory as found in \cite{Cheung:2018oki}, but beyond that, there are no new theories. Next, we couple vector fields to scalars and prove that there are no non-trivial higher derivative theories fixed by vanishing soft limits. Finally, we consider non-vanishing soft theorems and show that they might require the coupling of fields of various spins. We focus on the case of scalar-vector theories with fixed power-counting and demonstrate that there is a unique theory fixed by these soft theorems which we call \emph{Special scalar-vector Galileon} as the scalar part of this theory corresponds to the Special Galileon. We end with the discussion of more general vector Galileon theories which are related to theories studied in \cite{Bonifacio:2019hrj}.

\section{Soft bootstrap}
\label{sec:softbootstrap}

The low-energy effective field theories are defined by the Lagrangian which can have in principle an infinite tower of terms with undefined coupling constants. When the physical system exhibits certain symmetry, it can dramatically constrain the form of the Lagrangian and fix some (or all) coupling constants. The symmetries of the Lagrangian have profound consequences on the properties of the perturbative S-matrix, and in the low energy context can be often encoded in the \emph{soft limit} behavior. 

The \emph{soft bootstrap} is based on the alternative point of view. We start with the space of all tree-level S-matrices which have only physical poles and factorize properly. Such S-matrix \emph{ansatz} has a certain number of unfixed parameters which are in one-to-one correspondence (up to field redefinitions) to free coupling constants in Lagrangian (when redundant terms are eliminated by equations of motion and total derivatives). The idea is to impose a prescribed soft limit behavior, fix most (or all) parameters and specify a unique (or at least finite) set of special theories.

We start with certain lowest-order seed amplitudes (contact vertices), glue them together to ensure the right factorization, add all allowed contact terms multiplied by free constants and then fix these constants by demanding certain soft limit behavior of the full amplitude. We assume that the lowest-order seed vertices form a basis and similarly do the higher-order contact terms. If the soft bootstrap method gives us only a trivial solution (i.e. equivalent to free theory), we can claim that no theory with this soft limit behavior exists. In the case that we get non-trivial solutions we have to elaborate further. In particular, we have to check that such candidate corresponds to a self-consistent theory. This fact was in great detail discussed in \cite{Cheung:2016drk} and was motivated by the ``four-particle test'' \cite{Benincasa:2007xk}. 

In what follow we will concentrate on the case of bosonic fields only (for fermions in this context see e.g. \cite{Huang:2015sla,Elvang:2017mdq}). The important characteristics of the EFT Lagrangians is the number of derivatives. We can schematically write a term with $m$ derivatives on $n$ fields as
\begin{equation}
    {\cal L}_{m,n} = \partial^m \Phi^n\,,
\end{equation}
where $\Phi$ stands for any bosonic field. We will characterize a particular power-counting by  
\begin{equation}
\rho = \frac{m-2}{n-2}\,.
\end{equation}
The parameter $\rho$ defines an infinite tower of possible terms in the Lagrangian. For example, for scalar theories with $\rho=2$ we have schematically
\begin{align}
    {\cal L} &= \tfrac12 (\partial\phi)^2 + c_4(\partial^6\phi^4) + c_6 (\partial^{10}\phi^{6}) + c_8(\partial^{14}\phi^8)  + \dots \notag\\
    &= (\partial\phi)^2\bigl(\tfrac12 + c_4 (\partial^2\phi)^2 + c_6 (\partial^2\phi)^{4} + c_8 (\partial^2\phi)^6  + \dots\bigr)\,.\label{Lag}
\end{align}
In fact, at each order we get numerous different terms and multiple coupling constants $c_m \rightarrow c_m^{(p)}$. We see that the interpretation of $\rho$ is the average number of derivatives per one field $\phi$ once $(\partial\phi)^2$ is stripped off as evident from the second line in (\ref{Lag}).

The soft bootstrap approach is following: generate an ansatz for the tree-level amplitudes where each independent kinematical invariant is multiplied by a free parameter. These parameters are in one-to-one correspondence with the independent parameters in the Lagrangian (\ref{Lag}) after the off-shell redundancies are removed. It is straightforward to build an ansatz for kinematical terms, while removing redundancies from Lagrangian (total derivatives, equations of motion) is generally cumbersome.

In the next step, we impose additional kinematical constraint: the soft limit. We demand that in the soft limit of any external leg, i.e. for $p\to0$, the amplitude behaves as
\begin{equation}
A(p) = O(p^\sigma)\,,
\end{equation}
with the parameter $\sigma$ characterizing the soft limit scaling. The goal of the soft bootstrap is to explore the space of all theories (with various particle content) for fixed $(\rho,\sigma)$, find unique solutions or prove there are no such theories. 

Soft limits are directly linked to the symmetries of the Lagrangian (see \cite{Cheung:2016drk} for detailed discussion). This means that the amplitude with a special soft behavior indicates the existence of the theory with some special symmetries.

\medskip

For the scalar EFTs the classification has been done in \cite{Cheung:2016drk}. It is obvious that the theories with $\sigma<\rho$  are ``trivial'' solutions since the Lagrangian has at least $\sigma$ derivatives per field. The interesting solutions have $\sigma\geq \rho$ (except for $\rho=\sigma=1$ which is still trivial), i.e. the number of derivatives is smaller, allowing for intricate cancellations between Feynman diagrams in the soft limit caused by some underlying symmetries. 

The special scalar theories with 4pt fundamental vertices are:

\begin{enumerate}
    \item $(\rho,\sigma) = (0,1)$: Non-linear sigma model
    \item $(\rho,\sigma) = (1,2)$: Dirac-Born-Infeld action
    \item $(\rho,\sigma) = (2,2)$: Galileon theories
    \item $(\rho,\sigma) = (2,3)$: Special Galileon 
\end{enumerate}

All non-trivial theories with $\sigma>4$ are excluded. Some of these theories also appear in the context of color-kinematics duality and CHY formalism \cite{Cachazo:2014xea}. 

In what follows we will mainly concentrate on theories which are related to the Special Galileon \cite{Cheung:2014dqa,Cachazo:2014xea}. The latter  is a very interesting theory, since it occupies the very corner of the allowed region in the $(\rho,\sigma)$ plane. 
Its soft scaling $\sigma=3$ is the largest possible one among the nontrivial theories with $\sigma\geq\rho$.
On the Lagrangian level such an enhanced soft limit is a consequence of a polynomial shift symmetry \cite{Cheung:2016drk}. The particular form of this symmetry depends on the parametrization of the Lagrangian.
In the standard case we write the Lagrangian in $D$ dimensions in the form
\begin{equation}
\mathcal{L}_\mathrm{b}=\frac{(-1)^{D-1}}{D!}\sum_{n=0}^{D}d_{n+1}\varphi \varepsilon ^{\mu _{1}\ldots \mu
_{D}}\varepsilon ^{\nu _{1}\ldots \nu _{D}}\prod_{i=1}^{n}\partial _{\mu
_{i}}\partial _{\nu _{i}}\varphi \prod_{j=n+1}^{D}\eta _{\mu _{j}\nu _{j}},
\end{equation}
with $d_{2n+1}=0$ and with either
\begin{equation}
d_{2n}=\frac{(-1)^n}{2n}\left(\frac{1}{\alpha} \right)^{2(n-1)}\left( \begin{array}{c}
D \\ 
2n-1%
\end{array}%
\right), 
\end{equation}
or
\begin{equation}
d_{2n}=\frac{1}{2n}\left(\frac{1}{\alpha} \right)^{2(n-1)}\left( \begin{array}{c}
D \\ 
2n-1%
\end{array}%
\right), 
\end{equation}
where $\alpha$ is a dimensionful parameter which plays the role of the only coupling constant.
The symmetry transformation responsible for the soft behavior of the amplitudes is then given for the above two possibilities as \cite{Hinterbichler:2015pqa,Novotny:2016jkh}
\begin{equation}
\delta \varphi= G^{\mu\nu}(\alpha^2 x_\mu x_\nu \pm \partial_\mu\varphi\partial_\nu\varphi),
\end{equation}
where the parameter  $G^{\mu\nu}$ is an infinitesimal symmetric and traceless tensor. Here we formally distinguish two branches of the Special Galileon Lagrangian -- as we will see later, they can couple differently to other fields.

The uniqueness of the special theories mentioned above also allows us to reconstruct their amplitudes using \emph{soft recursion relations}. First  we have to set how exactly we will perform the single soft limit $p\to0$. It is clear that we cannot move only one momentum to zero without touching other momenta as they are related by the momentum conservation. In order to parametrize one momentum soft shift (for example as $p \to t p$) we need to touch also at least two other momenta. If we want to have simultaneous information on all possible soft limits in every line we can employ the so-called all-line shift \cite{Cheung:2015ota}. As we are working exclusively in $D=4$, it is indeed possible to define for the $n$-point kinematics with $n\geq6$ the following rescaling shift
\begin{equation}
    p_i \to p_i(1-z a_i)
\end{equation}
with all $a_i$ different. Then the studied amplitude has to behave as
\begin{equation}
    A_n(z) \sim (1-z a_i)^\sigma\qquad \text{for}\; z\to 1/a_i
\end{equation}
in all $i=1,\ldots n$ lines. In the theory with $\rho$ fixed, under the all-line shift the amplitude $A_n$ is expected to have the following  asymptotics for $z\to \infty$
\begin{equation}
    A_{n}(z)= O(z^{\rho(n-2)+2}).
    \label{asymptotics}
\end{equation}
In analogy with BCFW we can introduce a function
\begin{equation}
f_n(z)= \frac{A_n(z)}{F_n(z)}\,,
\end{equation}
where the denominator
\begin{equation}
F_n(z) = \prod_{i=1}^n (1-a_i z)^\sigma\,,
\end{equation}
is chosen so that the seeming poles $z=1/a_i$ cancel with the anticipated soft behavior of amplitude. Therefore the only singularities of $f_n(z)$ are the unitarity poles connected with the factorization channels of the amplitude $A_n$. Provided $\rho(n-2)+2<n\sigma$ we get
\begin{equation}
\label{eq:cauchy}
  f_n(z)\stackrel{z\to\infty}{=}0,\,\,\,\,\rightarrow\,\,\,\,\,\,\,\, \lim_{R\to\infty}\oint_{|z|=R} dz\frac{f_n(z)}{z}  = 0
  \end{equation}
and using the residue theorem for the meromorphic function $f_n(z)/z$
we can relate the amplitude $A(0)={\rm res}(f_n(z)/z,0) $ to the sum of residue of $f_n(z)/z$ at the unitarity poles.
Thus, in analogy with BCFW we can reconstruct the amplitude recursively from the sum of residua over the factorization channels.

\section{No-go theorems for EFTs}

In this section, we present a number of negative results in searching for new EFTs with vanishing soft limits. First, we explore the EFTs for spin-1 particles with $\rho=1,2$ and then coupled scalar-vector interactions with Galileon power-counting ($\rho=2$). In both cases we will assume for simplicity the conservation of parity. Let us also stress that all particles studied in this article are massless and thus we refer to spin-1 particle sometimes as vector and sometimes as photon but we mean the same thing.

\subsection{Interacting spin-1 theories}

Based on the successes of amplitude approach to uniquely fix certain scalar field theories we turn our attention to abelian massless vector fields. In four dimensions, there are two different helicity states $h=\pm 1$ and the amplitudes are characterized by the number of external particles $n$ and the number of negative helicity states $k$
\begin{equation}
    A_n(1^-2^-\dots k^-\,(k+1)^+\dots n^+)\equiv A_n(I^{+},I^{-})\,.
\end{equation}
While in the Yang-Mills theory the fundamental amplitudes are three-point and all higher helicity amplitudes for $k\in (2,\dots,n{-}2)$ are non-zero, in the context of higher derivative low-energy effective field theories it is natural to start with a four-point vertex. In order to ensure gauge invariance the Lagrangian must be a function of field strength $F^{\mu\nu}$. At the leading order in the derivative expansion, for $\rho=1$, the general Lagrangian can be written in a form
\begin{multline}
    {\cal L}^{\rho=1} = - \frac14 \langle FF\rangle + g_4^{(1)} \langle FFFF\rangle + g_4^{(2)} \langle FF\rangle^2 +\\+ g_6^{(1)} \langle FF\rangle^3 + g_6^{(2)} \langle FFFF\rangle\langle FF\rangle + g_6^{(3)} \langle FFFFFF\rangle + \dots\,, \label{Lagr1}
\end{multline}
where the traces $\langle\dots\rangle$ stand for $\langle FF\rangle = F^{\mu\nu}F_{\mu\nu}$, $\langle FFFF\rangle = F^{\mu\nu}F_{\nu\rho}F^{\rho\sigma}F_{\sigma\mu}$, etc. In four dimensions the Lagrangian ${\cal L}^{\rho=1}$ can be written in terms of two building blocks
\begin{equation}
    f\equiv -\frac14 F_{\mu\nu}F^{\mu\nu},\qquad g \equiv - \frac14 F_{\mu\nu}\widetilde{F}^{\mu\nu}\label{fg},
\end{equation}
where $\widetilde{F}^{\mu\nu}=\frac12\epsilon^{\mu\nu\rho\sigma}F_{\rho\sigma}$. The Lagrangian (\ref{Lagr1}) then becomes a polynomial in $f,g$,
\begin{equation}
    {\cal L}^{\rho=1} = f + a_1 f^2 + a_2 g^2 + b_1 f^3 + b_2 fg^2 + \dots\,, \label{Lagr2}
\end{equation}
where in parity conserving theories $g$ only enters in even powers. From this Lagrangian we can construct scattering amplitudes parametrized by coefficients $a_i$, $b_i$, etc. and ask if there are any constraints at the level of tree-level S-matrix which would fix these coefficients and specify a unique theory. 

The simple soft limit  $p\rightarrow0$ becomes richer in four dimensions when using spinor helicity formalism. The momentum $p^\mu$ factorizes into a product of two spinors\footnote{In the following sections we will use more convenient bracket notations for the spinors, e.g. $2p_1\cdot p_2=s_{12}=\langle12\rangle[21]$.} $p^\mu=\sigma^{\mu\dot{A}A} \lambda_A \widetilde{\lambda}_{\dot{A}}$ and we can define a \emph{chiral} soft limit when only one of the spinors is sent to zero, either holomorphic $\lambda\rightarrow0$ or anti-holomorphic $\widetilde{\lambda}\rightarrow0$. 
 We can then impose an enhanced soft limit behavior of the tree-level amplitude as a constraint,
\begin{equation}
    \lambda\rightarrow t\lambda,\quad \mbox{and}\qquad \lim_{t\rightarrow0}A_n = O(t^\sigma)\,.
\end{equation}
Here we assume that for helicity plus particles we choose the holomorphic (and for helicity minus the antiholomorphic) chiral soft limit in order  to avoid the artificially enhanced soft behavior caused by the deformation of the polarization spinors. As pointed out in \cite{Cheung:2018oki} there are no combinations of parameters such that the Lagrangian (\ref{Lagr2}) would exhibit an enhanced soft limit behavior, in this case $\sigma>0$. This could be achieved by performing multiple chiral soft limit when all holomorphic spinors of  helicity plus particles are sent to zero,
\begin{equation}
\lambda_i\rightarrow t\lambda_i,\,\,\mbox{for $i\in I^{+}$}\,,  \label{chiralS}
\end{equation}
while also shifting (some of) the antiholomorphic spinors $\widetilde{\lambda}_i\to\widetilde{\lambda}_i(t)$ for $i\in I^{-}$ of other particles to preserve the momentum conservation. The $n$-pt scattering amplitude $A_n$ derived from (\ref{Lagr2}) has generically $O(t^0)$ behavior for $t\rightarrow 0$ but imposing enhanced behavior,
\begin{equation}
   \lim_{t\rightarrow0}A_n = O(t^1)\,,
\end{equation}
we can fix all parameters (up to an overall scale $\Lambda$) and fix the amplitude, and hence also the Lagrangian (\ref{Lagr2}) uniquely. The resulting Lagrangian is 
\begin{equation}
    {\cal L}_{BI} = \Lambda^4-\Lambda^4\sqrt{1-2\frac{f}{\Lambda^4}-\frac{g^2}{\Lambda^8}}\,,
\end{equation}
which is the action for Born-Infeld (BI) theory\footnote{This action corresponds at the same time to the bosonic sector of the ${\cal N}=2$ SUSY Born-Infeld model. The soft behavior of the amplitudes can be then related to the ${\cal N}=2\to {\cal N}=1$ SUSY breaking in this model \cite{Cheung:2018oki}.}. Interestingly, all amplitudes in this theory preserve helicity, i.e. $k=n/2$ and we have the same number of plus and minus helicity states in any amplitude, but this condition itself is not enough to specify BI uniquely  (e.g. Bossard-Nicolai Lagrangian \cite{Bossard:2011ij} also preserves helicity while lacks any special soft limit behavior, see the detailed discussion in \cite{Novotny:2018iph}). The BI theory and the preservation of helicity has also been recently studied at one-loop \cite{Novotny:2018iph,Elvang:2019twd,Elvang:2020kuj}. 

Following the successful path of exploring scalar EFTs using soft limits, we can consider the Lagrangian with higher derivative terms and impose stronger behavior in the soft limit. In particular, we can shift various $\lambda$ and/or $\widetilde{\lambda}$ spinors for $+$ and/or $-$ helicity particles, send them to zero demanding
\begin{equation}
   \lim_{t\rightarrow0}A_n = O(t^\sigma)
\end{equation}
and fixing coefficients in the Lagrangian. The first case of interest is the Lagrangian generated by the $\partial^2F^4$ terms, i.e. for $\rho=2$. Adding the infinite tower of higher point contact terms, we can schematically write
\begin{equation}
    {\cal L}^{\rho=2} = -\tfrac14 F^2 + \sum_k c_4^{(k)} \partial^2F^4 + \sum_k c_6^{(k)} \partial^4F^6 + \sum_k c_8^{(k)} \partial^6 F^8 + \dots\,,
\end{equation}
where at each order we have to sum over all possible contractions of the Lorentz indices. We prefer to formulate this question alternatively in the language of tree-level amplitudes. Demanding that the helicity is preserved, we find one non-trivial 4pt amplitude,
\begin{equation}
    A_4(1^-2^-3^+4^+) = s_{12}\langle 12\rangle ^2[34]^2\quad\leftrightarrow \quad {\cal L}_4 = \partial f\cdot \partial f+\partial g\cdot \partial g\,,
\end{equation}
where $f$ and $g$ are the invariants (\ref{fg}). Here and in what follows, $s_{ijk\dots}=(p_i+p_j+p_k+\dots)^2$. Note that all other helicity preserving Lagrangian terms of the form $\partial^2F^4$ can be eliminated using equations of motion or integration by parts and thus do not contribute to 4pt amplitudes. At six-point, we are interested in the amplitude $A_6(1^-2^-3^-4^+5^+6^+)$. We have to sum over the factorization diagrams and the contact terms, schematically represented as
\vspace{4pt}
\begin{equation}
\begin{picture}(0,0)
\put(33,27){$1^-$}
\put(33,5){$2^-$}
\put(33,-27){$4^+$}
\put(87,5){$P^+$}
\put(142,5){$-P^-$}
\put(200,27){$3^-$}
\put(200,5){$5^+$}
\put(200,-27){$6^+$}
\end{picture}
{\rm i}A_6 = \vcenter{\hbox{\includegraphics[scale=0.73]{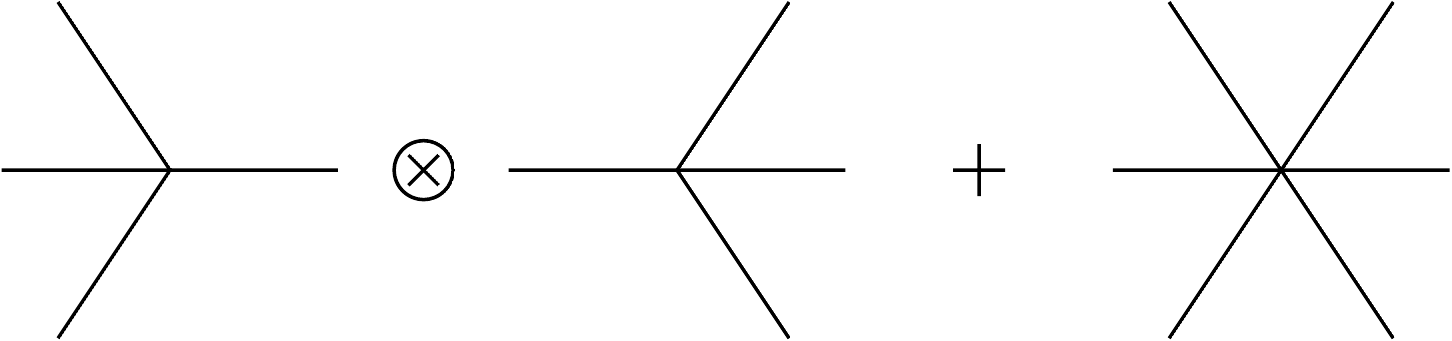}}}
\vspace{10pt}
\end{equation}
where all permutations are tacitly assumed.
The factorization diagrams can be obtained from on-shell gluing of 4pt amplitudes. For one factorization channel\footnote{Here we use the convention $|{-}P\rangle={\rm i}|P\rangle$.},
\begin{equation}
  -(s_{12}\la12\ra^2[4P]^2)\times\frac{1}{s_{124}}\times (s_{56}\la3\vert{-}P\ra^2[56]^2) \rightarrow \frac{\la12\ra^2[56]^2s_{12}s_{56}\la3|1{+}2|4]^2}{s_{124}} \,.\label{merge}
\end{equation}
Any difference between this prescription and another one -- where we represent the off-shell 4pt Feynman vertex in a different way -- would correspond to a contact term. Next, we generate an ansatz for all contact terms which have the correct derivative power-counting and little group weights. At 6pt, there are 5 independent kinematical terms. A particular choice of this basis is:
\begin{align}
    C_6 &= c_1\left(s_{12}\la23\ra^2\la13\ra^2[36]^2[45]^2\right) + c_2 \left(s_{12}\la12\ra^2\la13\ra\la23\ra[16][26][45]^2\right)\nonumber\\& \hspace{1cm}+ c_3 \left(s_{12}\la12\ra^2\la13\ra^2[16]^2[45]^2\right)+ c_4 \left(s_{14}\la12\ra^2\la13\ra\la23\ra[16][26][45]^2\right)\nonumber\\ & \hspace{1cm} + c_5\left(\la12\ra^2\la23\ra^2\la24\ra[14][16][26][45]^2\right) \,.\label{cont6}
\end{align}
All terms are then symmetrized in legs 1, 2, 3 and 4, 5, 6. The general ansatz for $A_6\equiv A_6(1^-2^-3^-4^+5^+6^+)$ is then
\begin{equation}
    A_6 = \sum_{\cal P} \frac{\la12\ra^2[56]^2s_{12}s_{56}\la3|1{+}2|4]^2}{s_{124}} + C_6(c_1,\dots,c_5)\,, \label{ans6}
\end{equation} 
where ${\cal P}$ denotes symmetrization. Following $\rho=1$ case we shift $A_6$ using (\ref{chiralS}) and impose soft behavior for $t\rightarrow0$, $A_6=O(t^\sigma)$ for certain $\sigma$. Note that we perform multi-chiral soft limit so the connection between $\rho$ and $\sigma$ is not that straightforward as in the single soft limit case. 

First, we observe that the factorization terms (when symmetrized) behave like $O(t^4)$, while the contact terms have general $O(t^2)$ behavior. There exists one linear combination of these terms which have enhanced $O(t^4)$ scaling,
\begin{equation}
c_2 = \frac{c_1}{2}, \quad c_3 = c_1, \quad c_4=c_5=0\,. \label{fix1}
\end{equation}
Plugging back in (\ref{cont6}) we get one-parametric contact term,
\begin{equation}
    C_6 = c_1\sum_{{\cal P}}s_{12}\la13\ra[45]^2\left( \la23\ra^2\la13\ra[36]^2+ \frac12 \la12\ra^2\la23\ra[16][26]+ \la12\ra^2\la13\ra[16]^2\right) \,,
\end{equation}
where ${\cal P}$ stands for the symmetrization in $1,2,3$ and $4,5,6$. There exists no value of coefficient $c_1$ such that the (\ref{ans6}) exhibits even stronger $O(t^5)$ scaling. As a result, we get one parametric solution and the theory is not unique. While there is a way how to fix this one parameter at six-point using a different multi-chiral soft limit (sending only two $\lambda$s to zero), this breaks again at 8pt. The ansatz for the 8pt contact term $C_8$ has 94 independent coefficients and the soft limit behavior fails to fix all of them leaving us with multiple free parameters.  
We performed various soft shifts including sending some momenta to zero at different rates, or introducing more parameters in the soft-limit, none of that produced a non-trivial answer.
Namely, for the 8pt amplitude we considered $\{a,b\}$ shift where we perform the soft shift of some number $a$ of $\lambda$ spinors and $b$ different $\widetilde{\lambda}$ spinors,
\begin{equation}
\lambda_{i_k} \rightarrow z\lambda_{i_k},\,\,\,k=1,2,\dots,a,\qquad \widetilde{\lambda}_{j_k} \rightarrow z\widetilde{\lambda}_{j_k},\,\,k=1,2,\dots b.
\end{equation}
In this notation, the standard holomorphic soft limit corresponds to $\{1,0\}$, while sending half of the $\lambda$s to zero would be $\{4,0\}$. Here we consider more general shift, but still we have not found any non-trivial soft limit scaling.
Furthermore, we generalized the shift by sending various spinors to zero at different rate. This was done just for holomorphic shifts of two, three or four $\lambda$ spinors. 
\begin{equation}
\lambda_{i_k} \rightarrow z^\# \lambda_{i_k},
\end{equation}
We tried $z$, $z^2$, $z^3$ powers in various combinations but none of that lead to any non-trivial soft limit behavior either. 
For example, using the three-particle soft limit $\{3,0\}$ and sending the spinors to zero at the same rate, we find multiple solutions at $O(t^2)$, and one contact term with $O(t^3)$ behavior. While this specifies a unique amplitude, it also eliminates all factorization channels leaving us just with a contact term. Other shifts (including shifting $\lambda,\widetilde{\lambda}$ for both $+$ and $-$ helicities) lead to the same conclusion: There is no non-trivial 8pt amplitude with special soft-limit behavior. This does not prove there is no way how to fix the amplitude uniquely, but at the moment we have to conclude that no standard soft limit checks work. This is consistent with the statements in the EFT literature that there is no ``vector-Galileon'' theory \cite{Deffayet:2013tca}. 

\subsection{Scalar-vector theories}
\label{sec:nogoscalphot}

Since we were unable to find a unique pure vector theory fixed by standard soft limits, we can try to couple vectors with other particles. The simplest choice is to couple vectors with scalars. There is some chance that this combined theory containing both scalars and vectors could be fixed uniquely via some non-trivial relations between the amplitudes. For simplicity, we concentrate only on single-$\rho$ theories with Galileon power-counting (i.e. $\rho=2$) which are parity conserving and which conserve helicity. We also suppose that their only non-zero amplitudes are with an even number of legs. As in the previous section, we formulate the problem using the tree-level amplitude language. 

By demanding $\rho=2$ and the correct little group scaling, we find there are three (contact) 4pt amplitudes possible for various combinations of scalars and vectors,
\begin{equation}\label{eq:nspt4pt}
\begin{split}
A_{04}(1^\varphi2^\varphi3^\varphi4^\varphi) &= c_{04} \, s_{12}s_{13}s_{23}\\
A_{22}(1^+2^-3^\varphi4^\varphi) &= c_{22}\, s_{34}[1|3|2\ra[1|4|2\ra\\
A_{40}(1^+2^+3^-4^-)&= c_{40}\, s_{12}[12]^2\la34\ra^2.
\end{split}
\end{equation}
The amplitudes are denoted $A_{n_\gamma n_\varphi}$, where $n_\gamma$ is the number of vector particles (or ``photons'' in what follows), $n_\varphi$ is the number of scalars, and the constants $c_{n_\gamma n_\varphi}$ are free couplings. Apart from the pure vector 4pt amplitude $A_{40}$, there are two additional 4pt amplitudes containing scalars. Just like in the previous section, any 6pt amplitude is constructed by gluing appropriate 4pt vertices together and adding independent 6pt on-shell contact terms with unknown constants. We can try imposing certain soft limits on the constructed ansatz to fix both the 4pt and 6pt constants.

There are four 6pt amplitudes: $A_{06}$, $A_{24}$, $A_{42}$, and $A_{60}$.
Both the pure scalar one $A_{06}$ and the pure vector one $A_{60}$ have been examined in the previous sections. The remaining two amplitudes $A_{24}$ and $A_{42}$ are new. The diagrams contributing to the amplitude $A_{24}$ are
\vspace{4pt}
\begin{equation}
{\rm i}A_{24} \,=\,  \vcenter{\hbox{\includegraphics[scale=0.73]{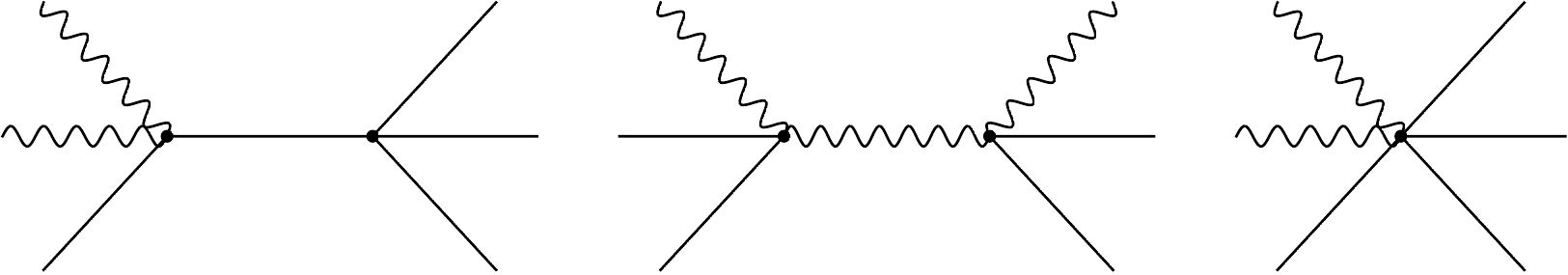}}}
\vspace{10pt}
\end{equation}
There are 29 independent 6pt on-shell contact terms, 
\begin{equation}
    v_6(1^+2^-3^\varphi4^\varphi5^\varphi6^\varphi) = \Big\{s_{12}^3 [1|3|2\ra^2, s_{34}^3 [1|3|2\ra^2, s_{56}^3 [1|3|2\ra^2, \dots\Big\} + {\rm perm}\,.
\end{equation}
Demanding the $O\left(t^3\right)$ soft limit behavior\footnote{This means we try to couple the photons to Special Galileon. This was suggested originally in \cite{,Elvang:2019twd} and also discussed in \cite{Bonifacio:2019rpv} from the Lagrangian point of view.} for any scalar leg fixes all these coefficients uniquely. This is in agreement with the reconstructibility condition (\ref{eq:cauchy}). After performing the all-line shift of the external momenta and dividing the deformed amplitude by the cubic soft terms $(1-a_iz)^3$, we find that $A_{n_\gamma n_\varphi}$ scales as
\begin{equation}
f_{n_\gamma n_\varphi}(z)=\frac{A_{n_\gamma n_\varphi}(z)}{\prod_{i=1}^{n_\varphi}(1-a_iz)^3} \sim \frac{z^{2(n_\gamma+n_\varphi)-2-n_\gamma}}{z^{3n_\varphi}}=O(z^{n_\gamma-n_\varphi-2}) \\,\,\, \mathrm{for}\,\, z\to\infty,
\label{e:cond}
\end{equation}
and thus $f_{n_\gamma n_\varphi}\to 0$  for $n_\gamma=2$ and $n_\varphi=4$. 
As discussed in the previous section, this means that $A_{24}$ is reconstructible.

The amplitude $A_{42}$ is more interesting. The contributing diagrams are
\vspace{4pt}
\begin{equation}
{\rm i}A_{42} \,=\,  \vcenter{\hbox{\includegraphics[scale=0.73]{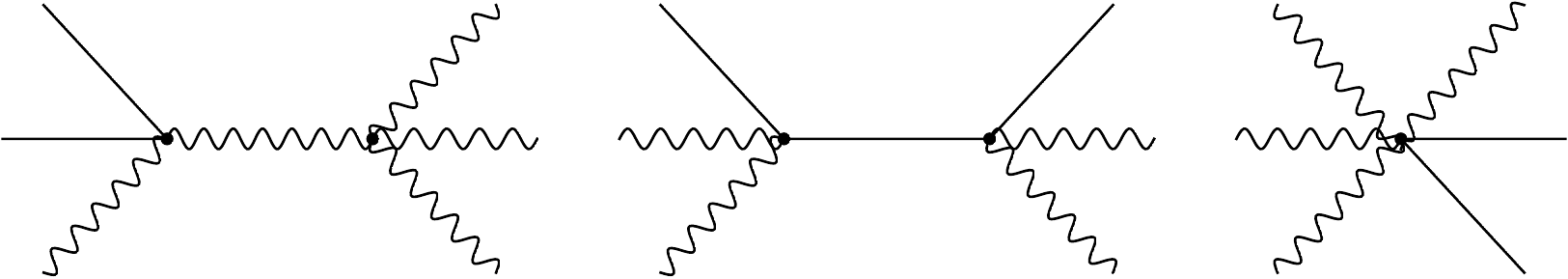}}}
\vspace{6pt}
\end{equation}
We have verified that demanding just the $O\left(t^3\right)$ soft limit behavior in any scalar leg is not enough to fix the amplitude uniquely. This again agrees with the considerations based on (\ref{e:cond}), for $n_\gamma=4$ and $n_\varphi=2$. Nevertheless, it is possible to fix the amplitude using an additional soft limit demanding the $O(t)$ multi-chiral soft limit in two of the photonic legs. Interestingly, this soft limit also fixes one of the 4pt constants, either $c_{40}$ or $c_{22}$. Note that the purely photonic amplitude $A_{60}$ was also uniquely fixed using the multi-chiral soft limit behavior in two legs.

At 8pt we have five distinct amplitudes: $A_{08}$, $A_{26}$, $A_{44}$, $A_{62}$ and $A_{80}$. The pure scalar amplitude $A_{08}$ can be fixed by demanding the $O(p^3)$ soft behavior giving us the amplitude in the Special Galileon theory. The same procedure now allows us to fix the amplitude $A_{26}$ again using the soft limit in one of the scalar legs in agreement with the soft reconstructibility argument (\ref{e:cond}). However, all remaining amplitudes $A_{44}$, $A_{62}$ can not be possibly fixed using any standard soft limits. We have tried various combinations of soft limit vanishing in scalar legs and multi-chiral soft limits in photons, but nothing lead to a unique theory. This is consistent with our inability to fix the pure photonic amplitude $A_{80}$ in the previous subsection. 

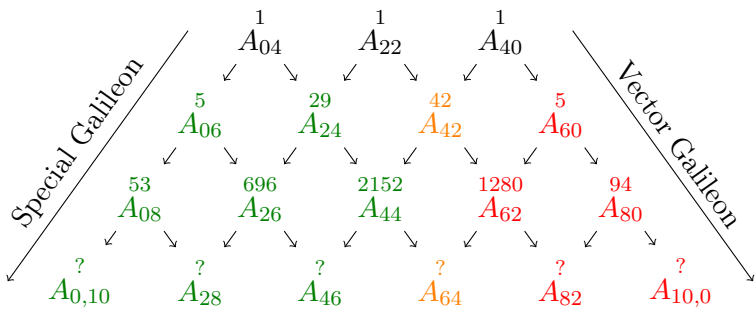
\begin{figure}[ht]
\begin{center}
\begin{tikzpicture}[x=50pt, y=35pt, scale=.9]
\def\dgreen{green!50!black}
\node (04) at (-1,3) {$\stackrel{1}{A_{04}}$};
\node (22) at (0,3) {$\stackrel{1}{A_{22}}$};
\node (40) at (1,3) {$\stackrel{1}{A_{40}}$};

\node (06) [text=\dgreen] at (-1.5,2) {$\stackrel{5}{A_{06}}$};
\node (24) [text=\dgreen] at (-.5,2) {$\stackrel{29}{A_{24}}$};
\node (42) [text=orange] at (.5,2) {$\stackrel{42}{A_{42}}$};
\node (60) [text=red] at (1.5,2) {$\stackrel{5}{A_{60}}$};

\node (08) [text=\dgreen] at (-2,1) {$\stackrel{53}{A_{08}}$};
\node (26) [text=\dgreen] at (-1,1) {$\stackrel{696}{A_{26}}$};
\node (44) [text=\dgreen] at (0,1) {$\stackrel{2152}{A_{44}}$};
\node (62) [text=red] at (1,1) {$\stackrel{1280}{A_{62}}$};
\node (80) [text=red] at (2,1) {$\stackrel{94}{A_{80}}$};

\node (010) [text=\dgreen] at (-2.5,0) {$\stackrel{?}{A_{0,10}}$};
\node (28) [text=\dgreen] at (-1.5,0) {$\stackrel{?}{A_{28}}$};
\node (46) [text=\dgreen] at (-.5,0) {$\stackrel{?}{A_{46}}$};
\node (64) [text=orange] at (.5,0) {$\stackrel{?}{A_{64}}$};
\node (82) [text=red] at (1.5,0) {$\stackrel{?}{A_{82}}$};
\node (100) [text=red] at (2.5,0) {$\stackrel{?}{A_{10,0}}$};

\draw [->] (04) -- (06);
\draw [->] (04) -- (24);
\draw [->] (22) -- (24);
\draw [->] (22) -- (42);
\draw [->] (40) -- (42);
\draw [->] (40) -- (60);

\draw [->] (06) -- (08);
\draw [->] (06) -- (26);
\draw [->] (24) -- (26);
\draw [->] (24) -- (44);
\draw [->] (42) -- (44);
\draw [->] (42) -- (62);
\draw [->] (60) -- (62);
\draw [->] (60) -- (80);

\draw [->] (08) -- (010);
\draw [->] (08) -- (28);
\draw [->] (26) -- (28);
\draw [->] (26) -- (46);
\draw [->] (44) -- (46);
\draw [->] (44) -- (64);
\draw [->] (62) -- (64);
\draw [->] (62) -- (82);
\draw [->] (80) -- (82);
\draw [->] (80) -- (100);

\draw [->] (-1.6,3) -- (-3.1,0) node[midway,sloped,above] {Special Galileon};
\draw [->] (1.6,3) -- (3.1,0) node[midway,sloped,above] {Vector Galileon};
\end{tikzpicture}
\end{center}
\vspace{-10pt}
\caption{The web of the amplitudes and the contact terms. We use the notation $A_{n_\gamma n_\varphi}^\#$, where $A_{n_\gamma n_\varphi}$ is either the set of contact terms (\# is their number), or the amplitude. Whenever two nodes of the web can be connected by an oriented path, the contact terms corresponding to the starting point of the path contribute to the amplitude attached to the endpoint of the path. Provided the theory exists, the green-colored amplitudes can be uniquely reconstructed recursively from the $O(p^3)$ soft limit alone, the red-colored ones should be fixed by some additional requirements. The orange-colored amplitudes $A_{n_\varphi+2,n_\varphi}$ could be reconstructed using just one such extra requirement.}
\label{f:web}
\end{figure}

To conclude, we have not been able to find a theory with Galileon power-counting, containing both scalars and vectors, uniquely fixed by the $O(p^3)$ soft limit and some type of multi-chiral limit(s). The results are summarized in Figure \ref{f:web}.

\section{EFTs with non-trivial soft theorem\label{soft theorems}}

As we have concluded in the previous two sections, apart from the Born-Infeld action there are no vector or scalar-vector EFTs which can be fixed by vanishing soft limit behavior. In this section, we consider more general type of the tree-level soft theorems,  schematically
\begin{equation}
    A_{n{+}1}(p,1,\dots,n)\stackrel{p\to 0}{=} \sum_i S_i(p) \times A_n^{(i)}(1,\dots,n) +O(p^{\sigma+1})
    \label{soft theorem schematically}
\end{equation}
when the leading $O(p^\sigma)$ term in the soft limit of the amplitude $A_{n{+}1}$ is equal to the linear combination of lower point amplitudes $A_n^{(i)}$ depending on the hard momenta $\{1,\dots,n \}$ multiplied by certain rational soft factors $S_i(p)$. The latter depends on both the soft and hard momenta. The amplitudes $A_n^{(i)}(1,\dots,n)$ on the right hand of (\ref{soft theorem schematically}) side can have completely different particle content in comparison with the amplitude $A_{n+1}$ on the left hand side, i.e. the helicities and the other discrete quantum numbers of the particles with momenta $\{1,\dots,n \}$ can be changed.
The Lorentz invariance, the little group scaling and the mass dimension of the amplitudes involved in (\ref{soft theorem schematically}) set constraints on the form of the soft factor $S_i(p)$ as well as on the soft exponent $\sigma$ and on the changes of the helicities of the particles in $A^{(i)}_n$.
For instance, as we will prove in what follows, in the case of the pure massless vector theories with $\rho=2$ studied in the previous sections,
and under some additional assumptions\footnote{It suffices to require that the soft factor does not depend on the reference spinors defining the polarization vectors of the soft particle.}, the only possible rational soft factor is $S_i(p)=0$. In order to get nontrivial soft factor in theories containing massless vectors it is therefore necessary to have also particles with other helicities in the spectrum of asymptotic states. The simplest possibility how to enlarge the vector theory is to add additional massless scalars. 

We will later derive the soft theorems in the massless single-flavor scalar-vector theory with $\rho=2$ power-counting, in the following form: 

\smallskip

\noindent $\bullet$ For soft positive helicity massless vector particle $h=+1$, 
\begin{eqnarray}
A_{n+1}\left( p^{+},I^{+},I^{-},I^{\varphi }\right) &\stackrel{p\to 0}{=}&%
\frac{a}{\Lambda^3}\sum\limits_{i\in I^{+}}\left[ p,i\right] ^{2}A_{n}\left(
I^{+},I^{-},I^{\varphi }\right) _{i^{+}\rightarrow i^{\varphi }}\notag\\
&&+\frac{b}{%
\Lambda^3 }\sum\limits_{i\in I^{\varphi }}\left[ p,i\right] ^{2}A_{n}\left(
I^{+},I^{-},I^{\varphi }\right) _{i^{\varphi }\rightarrow i^{-}}+O(p)\,,
\label{general soft photon theorem}
\end{eqnarray}
where $\Lambda$ is some scale and $a$ and $b$ are dimensionless parameters. We denoted $p^+$ the soft momentum of the positive helicity vector particle, $I^+$, $I^-$, $I^\varphi$ is a collective notation for positive or negative helicity vectors and scalars. The subscript of the amplitudes on the right hand side, $i^+\rightarrow i^\varphi$ and $i^\varphi\rightarrow i^-$, indicates the change of the type of particles corresponding to the momentum $i$. The soft theorem for particle with helicity $h=-1$ is analogous and within parity conserving theories it is parametrized by the same $a$, $b$ and $\Lambda$.

\smallskip

\noindent $\bullet$ For soft massless scalar particle,
\begin{eqnarray}
A_{n+1}\left( p^{\varphi},I^{+},I^{-},I^{\varphi }\right) &\stackrel{p\to 0}{=}&%
\frac{c}{\Lambda^3}\sum\limits_{i\in I^{+}\cup I^{-}}( p\cdot i)A_{n}\left(
I^{+},I^{-},I^{\varphi }\right) \notag\\
&&+\frac{d}{%
\Lambda^3 }\sum\limits_{i\in I^{\varphi }}( p\cdot i)A_{n}\left(
I^{+},I^{-},I^{\varphi }\right) +O(p^2).\label{general soft scalar theorem}
\end{eqnarray}

Now we will present a general discussion of the soft theorems (\ref{soft theorem schematically}) with various examples for different power-countings and helicities of the soft particles.

\subsection{General multi-spin soft theorems in EFTs}

Let us consider the scattering amplitude $A_{n{+}1}$ which generically behaves as $O(p^\sigma)$ in the soft limit due to power-counting. The enhanced  $O(p^{\sigma+1})$ behavior would correspond to a non-trivial, perhaps unique theory if exists, but suppose there is no such theory and instead we assume a generalized tree-level soft theorem in the  form (\ref{soft theorem schematically}), 
or more explicitly
\begin{equation}
A_{n+1}(p_{\alpha }^{h},1,2,\ldots ,n|\mathbf{h},\boldsymbol{\alpha })\overset{%
p\rightarrow 0}{=}\sum\limits_{i}S_{i}\left( p\right) A_{n}^{\left(
i\right) }\left( 1,2,\ldots ,n|\mathbf{h}^{\left( i\right) },\boldsymbol{\alpha }%
^{\left( i\right) }\right) +O\left( p^{\sigma +1}\right).
  \label{soft_theorem_ansatz}
\end{equation}
where $h,\alpha$ are the helicity and the discrete quantum numbers of the soft particle, while ${\bf{h}}=\left\{ h_{1},\ldots ,h_{n}\right\} $ denotes the helicities and
$\boldsymbol{\alpha} =\left\{ \alpha _{1},\ldots ,\alpha _{n}\right\}$
 the discrete quantum numbers of the hard particles in amplitude $A_{n+1}$.
The right-hand side includes $n$-pt amplitudes $A_n^{(i)}$ where the helicities ${\bf{h}}^{(i)}$ and quantum numbers $\boldsymbol{\alpha}^{(i)}$ are generally different form ${\bf{h}}$ and $\boldsymbol{\alpha}$.

In the above ansatz,
 $S_i(p)\equiv S_i(|p\rangle,|p],\{|j\rangle,|j]\}_{j=1}^{n})$ is an appropriate rational soft factor built from spinors corresponding to the momenta $\{p,1,2,\dots,n\}$.
 Note that in the case of soft photon, gluon or graviton, the soft factor might also depend on the reference spinors defining the polarization vectors. Since we assume only the single-$\rho$ higher derivative EFTs with interaction Lagrangian built solely in terms of $F_{\mu\nu}$ or  linearized Riemann tensor, we do not consider this possibility in what follows. 
 We further assume $S_{i}(p)$ to be a homogeneous function of  all its arguments and also separately as a function of  the spinors $|p\rangle,|p]$ only.
Let us define the holomorphic and antiholomorphic weights $a_i$ and $s_i$ according to
\begin{equation}
  S_i(t|p\rangle,u|p],\{t|j\rangle,u|j]\}_{j=1}^{n})=t^{a_i}u^{s_i} S_i(|p\rangle,|p],\{|j\rangle,|j]\}_{j=1}^{n}) 
\end{equation}
and similarly the weights $a^p_i$ and $s^p_i$ as
\begin{equation}
  S_i(t|p\rangle,u|p],\{|j\rangle,|j]\}_{j=1}^{n})=t^{a^p_i}u^{s^p_i} S_i(|p\rangle,|p],\{|j\rangle,|j]\}_{j=1}^{n}). 
\end{equation}
In the single-$\rho$ theory we have the following balance of the degrees of homogeneity of both sides of the  relation (\ref{soft_theorem_ansatz})
\begin{equation}
  \rho(n+1-2)+2=\frac{1}{2}(a_i+s_i)+\rho(n-2)+2,  
\end{equation}
while the right little group scaling requires
\begin{equation}
2 h+2|{\bf{h}}| =   s_i-a_i+2|{\bf{h}}^{(i)}|,
\end{equation} 
where we have denoted $|{\bf{h}}|\equiv \sum_{i=j}^{n}h_j$ the total helicity of the hard  particles in the amplitude $A_{n+1}$ and similarly $|{\bf{h}}^{(i)}|$ is the total helicity of the particles involved in the amplitude $A_n^{(i)}$.
Also, using the little group scaling for the soft particle and the required $O(p^\sigma)$ soft scaling\footnote{Note that taking the $p\to 0$ limit we do not deform the holomorphic spinors for  $h<0$ and the antiholomorphic spinors for $h>0$.}, 
\begin{equation}
2h=s^p_i-a^p_i, \,\,\,\,2\sigma+2|h|=s^p_i+a^p_i  \,,
\end{equation}
where $a^p_i$ and $s^p_i$ is the holomorphic and antiholomorphic weights of $S_i(p)$ with respect to the spinors $|p\rangle$ and $|p]$ spinors only.
The solution of these constraints reads\footnote{Note that these constraints can be easily adopted for multi-$\rho$ theories when the amplitudes $A_{n+1}$ and $A_n$ are characterized by different $\rho$'s. In such a case we have to replace $\rho\to\Delta\rho$, where $\Delta\rho=\rho_{n+1}(n-1)-\rho_n(n-2)$ and where the  $\rho$ parameters of the amplitude $A_{n+1}$ and $A_n$ are $\rho_{n+1}$ and $\rho_n$ respectively.}
\begin{eqnarray}
a_i&=&\rho-h-\Delta_i|{\bf{h}}| ,\,\,\,\,
s_i=\rho+h+\Delta_i|{\bf{h}}|\notag\\
a^p_i&=&\sigma+|h|-h,\,\,\,\,\,\,\,\,\,\,\,\,\,\,\,s^p_i=\sigma+|h|+h\,,
\label{general soft theorem constraints}
\end{eqnarray}
where $\Delta_i|{\bf{h}}|\equiv |{\bf{h}}|-|{\bf{h}}^{(i)}| $ is the change of the total helicity of hard particles.

These general constraints can be further simplified under the assumption that in the $i-$th term on the right hand side of (\ref{soft_theorem_ansatz}), the soft factor $S_i(p)$ depends only on $|p\rangle$, $|p]$ and  the spinors $|i\rangle$, $|i]$ corresponding to the $i-$th particle, and that  only the $i-$th particle changes its quantum numbers, i.e. 
 $ h_j^{(i)}=h_j$ and $\alpha_j^{(i)}=\alpha_j$ for $j\neq i$.
 These assumptions are motivated by the most usual mechanism of appearance of such soft factors. They are typically caused by Feynman diagrams with soft particle attached to $i-$th hard external line via three-point vertex\footnote{Note, however, that the presence of three-point vertices is not the only source of nontrivial soft factors. Explicit example of more complex mechanism can be found in \cite{Kampf:2019mcd}.
 }. The form of the soft factor and the helicity change are then generated as a result of interplay between the soft limit of the three-point vertex  and the off-shell propagator connecting it with the rest of the graph. 
 Both these building blocks depend just on $|p\rangle$, $|p]$ and $|i\rangle$, $|i]$ and only the quantum numbers of $i-$the particle can be changed.
 For particular realization of such a mechanism see Appendix \ref{soft_theorems}.
 
Under the above assumptions, the only building blocks for $S_i(p)$ are then the brackets $\langle p|i\rangle$ and $[p|i]$ and thus $a_i=2a_i^p$,  $s_i=2s_i^p$. The general form of the soft theorem reads
\begin{equation}
A_{n+1}(p_{\alpha }^{h},1,2,\ldots ,n|\mathbf{h},\boldsymbol{\alpha })\stackrel{p\to 0}{=}\sum_{i=1}^{n}
  C_i\langle p|i\rangle^{a_i^p} [p|i]^{s_i^p}
  A_n(1,2,\ldots ,n|\mathbf{h}^{(i)},\boldsymbol{\alpha }^{(i)})+O(p^{\sigma+1}),
\end{equation}
where $C_i$ are some constants and where now
\begin{equation}
 a_i^p=\frac{1}{2}\rho-h   ,\,\,\,\,s_i^p=\frac{1}{2}\rho+h,\,\,\,\,\Delta_i |{\bf{h}}|=h,\,\,\,\,\sigma=\frac{1}{2}\rho-|h|,
\end{equation}
i.e. on the right hand side, the helicities of the hard particles are given in terms of the helicities on the left hand side as
\begin{equation}
{\bf{h}}^{(i)}=\{h_1,h_2,\dots,h_i-h,\dots,h_n\}.
\end{equation}
Note that the weights $a_i^p$, $s^p_i$ and $\sigma$ have to be integers, which excludes some of the  pairs $(\rho, h)$.

\subsection{Examples of multi-spin soft theorems}

We will now apply the general soft theorems derived in the previous section to  particular cases of interest: $\rho=0,1,2$ power-counting.

\smallskip

\noindent $\bullet$ \underline{$\rho=0$}: The only possible case of a nontrivial soft theorem is for integer helicity $h=0,\pm1,\dots$ and the non-positive value $\sigma=0,-1,-2,\dots$. For a scalar theory, $h=0$, we get
\begin{equation}
A_{n+1}(p_{\alpha },1,2,\ldots ,n|\boldsymbol{\alpha })\stackrel{p\to 0}{=}\sum_{i=1}^{n}
  C_i
  A_n(1,2,\ldots ,n|\boldsymbol{\alpha }^{(i)})+O(p).
\end{equation}
This type of the soft theorem was discussed in \cite{Kampf:2019mcd} in the context of nonlinear sigma models and in~\cite{Luo:2015tat} in the context of dilaton and non-Goldstone particles. In theories with massless vectors and for soft particle with helicity $h=\pm 1$ we get
\begin{equation}
A_{n+1}(p_{\alpha }^h,1,2,\ldots ,n|{\bf{h}},\boldsymbol{\alpha })\stackrel{p\to 0}{=}\sum_{i=1}^{n}
  C_i\frac{[p|i]^{h}}{\langle p|i\rangle^{h}}
  A_n(1,2,\ldots ,n|{\bf{h}},\boldsymbol{\alpha }^{(i)})|_{h_i\to h_i-h}+O(1).
\end{equation}
This form of the soft vector theorem case  was introduced in \cite{Elvang:2016qvq} as a new possible subleading term in soft photon theorem within higher derivative EFTs.

A combination of both soft theorems can be studied in theories that non-trivially mix scalar and spin-1 degrees of freedom. Typical example would be ${\cal N}=2$ $\mathbb{CP}^1$ NLSM, an important case in this context \cite{Elvang:2018dco}, which we leave open for future work. 
\smallskip

\noindent $\bullet$ \underline{$\rho=1$}: The only nontrivial soft factor is possible for half-integer helicities of the soft particle. For $h=1/2$ we get
\begin{equation}
A_{n+1}(p_{\alpha }^{1/2},1,2,\ldots ,n|{\bf{h}},\boldsymbol{\alpha })\stackrel{p\to 0}{=}\sum_{i=1}^{n}
  C_i[p|i]
  A_n(1,2,\ldots ,n|{\bf{h}},\boldsymbol{\alpha }^{(i)})|_{h_i\to h_i-1/2}+O(p).
\end{equation}
Similarly to previous case we leave a deeper study of possible realization of this soft theorem open for future (tentative model should non-trivially mix between bosons and spin-1/2). 

The soft factor for $h=3/2$ reads:
\begin{equation}
A_{n+1}(p_{\alpha }^{3/2},1,2,\ldots ,n|{\bf{h}},\boldsymbol{\alpha })\stackrel{p\to 0}{=}\sum_{i=1}^{n}
  C_i\frac{[p|i]^{2}}{\langle p|i\rangle}
  A_n(1,2,\ldots ,n|{\bf{h}},\boldsymbol{\alpha }^{(i)})|_{h_i\to h_i-3/2}+O(1).
\end{equation}

\noindent $\bullet$ \underline{$\rho=2$}: only integer $h$'s are possible for nontrivial soft factor and then
\begin{equation}
A_{n+1}(p_{\alpha }^h,1,2,\ldots ,n|{\bf{h}},\boldsymbol{\alpha })\stackrel{p\to 0}{=}\sum_{i=1}^{n}
  C_i\frac{[p|i]^{h}}{\langle p|i\rangle^{h}}(p\cdot i)
  A_n(1,2,\ldots ,n|{\bf{h}},\boldsymbol{\alpha }^{(i)})|_{h_i\to h_i-h}+O(p^{2-|h|}).
  \label{rho2 soft theorem}
\end{equation}
The case of scalar theory, i.e. $h=h_j=0$ was discussed in \cite{Kampf:2020tne} in connection with multi-flavor Galileon theories while the case $h=2$ was first established in \cite{Elvang:2016qvq} as a new subleading term in the soft graviton theorem in effective field theories.

 Let us now return to the general form (\ref{soft_theorem_ansatz}) of the soft theorem  and apply the constraints (\ref{general soft theorem constraints}) on the case we have discussed in the previous sections, namely to the  scalar-vector theory with Galileon power-counting.
For $h=1$ and  $\rho=2$ we get from (\ref{general soft theorem constraints})
\begin{eqnarray}
 a_i&=&1-  \Delta_i |{\bf{h}}|,\,\,\,\,s_i=3+  \Delta_i |{\bf{h}}| ,\notag\\
 a^p_i&=&\sigma,\,\,\,\,s^p_i=\sigma+2\,.
\end{eqnarray}
Let us first assume a theory which contains only massless vector particles in the spectrum of asymptotic states.
Since  both  $a_i$ and $s_i$ have to be even  due to the Lorentz invariance and since for  $h_j=\pm 1$ we get only even helicity change  $\Delta h_i=0,\pm 2$, the only consistent possibility in pure vector theory is $S_i(p)=0$.
Therefore, to get nontrivial soft factor for the case of  single soft particle with $h=\pm1$ we need to add  other particles with different spins to the theory which could produce odd sum of the helicity changes.
The simplest possibility is to add scalars to the spectrum and, as we discussed above, to change the helicity only of the $i-$th particle in the $i-$th term of the sum on the right hand side of (\ref{soft_theorem_ansatz}). We assume that $S_i(p)$ depends only on $|p\rangle$, $|p]$ and  the spinors corresponding to the $i-$th particle. 
These assumptions lead to the soft theorems of the  general form (\ref{general soft photon theorem}) and (\ref{general soft scalar theorem}) mentioned above.

\section{Non-trivial soft bootstrap of scalar-vector Galileon}

We will now employ the non-trivial soft theorems in the bootstrap of the scalar-vector theory with Galileon power-counting $\rho=2$. 

\subsection{The simplest extension}

First, we will concentrate on the special form of the scalar soft limit, namely $O(p^3)$ behavior. It is the first natural generalization of completely vanishing soft limits, as studied in Section~\ref{sec:nogoscalphot}. We will thus demand here:
\begin{eqnarray}\label{eq:ansatzphotons}
A_{n+1}\left( p^{+},I^{+},I^{-},I^{\varphi }\right) &\stackrel{p\to 0}{=}&%
\frac{a}{\Lambda^3}\sum\limits_{i\in I^{+}}\left[ p,i\right] ^{2}A_{n}\left(
I^{+},I^{-},I^{\varphi }\right) _{i^{+}\rightarrow i^{\varphi }}\notag\\
&&+\frac{b}{%
\Lambda^3 }\sum\limits_{i\in I^{\varphi }}\left[ p,i\right] ^{2}A_{n}\left(
I^{+},I^{-},I^{\varphi }\right) _{i^{\varphi }\rightarrow i^{-}}+O(p)
\end{eqnarray}
for a plus helicity and similarly for a minus helicity and for scalars:
\begin{equation}\label{eq:ansatzop3}
A_{n+1}\left( p^{\varphi},I^{+},I^{-},I^{\varphi }\right) \stackrel{p\to 0} = O(p^3).
\end{equation}
This is a starting point for the soft amplitude  bootstrap. 
As before, we  generate the most general kinematical ansatz for the amplitude $A_{n{+}1}$ and then
fix the free parameters demanding the validity  of the above soft theorems. In this case, in addition to free parameters in the kinematical ansatz we also have additional parameters $a,b$ from the soft theorems.

As before we assume the parity conservation. Again, due to the Lorentz invariance for our power-counting only the amplitudes with  even number of vector particles are allowed. One important difference from the bootstrap setup in previous section, is the presence of odd amplitudes, and we need to consider vertices of any multiplicity in our ansatz. There are no 3pt amplitudes for $\rho=2$ power-counting, so the pure 4pt and 5pt vertices are immediately the corresponding amplitudes (i.e. they don't factorize to subamplitudes). By the nature of the vector soft theorem we cannot limit ourselves to helicity conserving amplitudes only. The 4pt amplitudes given already in (\ref{eq:nspt4pt}) must be thus extended by the  helicity-violating 4pt vertices to the following seven-parametric set
\begin{equation}\label{eq:4ptinputs}
\begin{split}
A_{04}(1^\varphi2^\varphi3^\varphi4^\varphi) &= c_{04} \, s_{12}s_{13}s_{23}\\
A_{22}(1^+2^-3^\varphi4^\varphi) &= c_{22}\, s_{34}[1|3|2\ra[1|4|2\ra\\
A_{22}(1^+2^+3^\varphi4^\varphi) &=  c_{22}'\, (s_{13} s_{23}+s_{14} s_{24})[12]^2 + c_{22}''\, s_{12}^2 [12]^2\\
A_{40}(1^+2^+3^-4^-)&= c_{40}\, s_{12}[12]^2\la34\ra^2\\
A_{40}(1^+2^+3^+4^+)&= c_{40}'\, \bigl(s_{12}[12]^2[34]^2 + s_{13}[13]^2[24]^2 + s_{23} [23]^2[14]^2\bigr)\\
A_{40}(1^+2^+3^+4^-)&= c_{40}''\, \bigl(\la14\ra^2[12]^2[13]^2 + \la34\ra^2[23]^2[13]^2+\la24\ra^2[12]^2[23]^2\bigr)\,.
\end{split}
\end{equation}
We can now proceed to the 5pt and 6pt vertices.  For reference purposes  let us summarize  the numbers of terms of the vertex basis. 
For the 5pt vertices there are in total 26 constants, schematically represented by the helicity state ($\pm$) for a photon and $\varphi$ for a scalar:
\begin{align}\label{eq:5ptbaze}
&{+}{+}{+}{+}\varphi\, (3),\; {+}{+}{+}{-}\varphi\, (4),\; {+}{+}{-}{-}\varphi\, (6),\;{+}{+}\varphi\varphi\varphi\, (6),\; {+}{-}\varphi\varphi\varphi\, (5),\;  \varphi\varphi\varphi\varphi\varphi\, (2)
\end{align}
(with an exact number of terms for the individual combination in parenthesis).
The number of terms grows rapidly, so for the 6pt vertices there are already 215 monomials:
\begin{align}
&{+}{+}{+}{+}{+}{+}\, (3),\; {+}{+}{+}{+}{+}{-}\, (4),\; {+}{+}{+}{+}{-}{-}\, (12),\; {+}{+}{+}{-}{-}{-}\, (5),\;{+}{+}{+}{+}\varphi\varphi\, (23),\;\notag\\
&{+}{+}{+}{-}\varphi\varphi\, (56),\; {+}{+}{-}{-}\varphi\varphi\, (42),\;{+}{+}\varphi\varphi\varphi\varphi\, (36),\;{+}{-}\varphi\varphi\varphi\varphi\, (29),\; 
\varphi\varphi\varphi\varphi\varphi\varphi\, (5)\,.
\end{align}
Requiring the validity of the soft theorems~(\ref{eq:ansatzphotons}) and~(\ref{eq:ansatzop3}) lead to a set of non-trivial relations entangling the 4pt, 5pt and the 6pt constants. The system is easy to reduce and the result is very surprising: the whole theory is governed by only one free parameter! So not only all the 5pt and 6pt vertices are completely fixed by the 4pt inputs (\ref{eq:4ptinputs}), but these 4pt inputs are subject of such strong conditions that they are all fixed. The same is true for the  parameters of the soft theorem, namely $a$ and $b$. Let us summarize the results. The parameters $a$ and $b$ are related as  
\begin{equation}\label{eq:ab}
    -a=b=\frac{1}{\Lambda^3},\,
\end{equation}
where $\Lambda$ is some dimensionful scale, and the solution for the 4pt vertices reads:
\begin{equation}\label{eq:sol4pt}
c_{04} = c_{22} = \frac{1}{\Lambda^6},\;
c_{40} = c_{40}'= -\frac{1}{\Lambda^6},\;
c_{22}'=c_{22}''=c_{40}''=0  \,.  
\end{equation}
Note that the  nontrivial amplitudes $A_{04}$ and $A_{22} $ obey the $O(p^3)$ enhanced Adler zero due to the momentum conservation, e.g.
\begin{equation}
A_{22}=\frac{1}{\Lambda ^{6}}s_{12} [3|2|4\rangle \lbrack
3|1|4\rangle =-\frac{1}{\Lambda ^{6}}s_{12} [3|2|4\rangle
^{2}=-\frac{1}{\Lambda ^{6}}s_{12} [3|1|4\rangle ^{2},
\end{equation}
and the amplitudes $A_{22}$ and $A_{40}$ vanish in the single soft vector
limit in accord with the soft theorem, since there are no 3pt on-shell amplitudes.

It is interesting that almost all 5pt amplitudes vanish. The only non-vanishing combination is $A_{41}(1^+2^+3^+4^-5^\varphi)$ and its parity conjugated one $A_{41}(1^-2^-3^-4^+5^\varphi)$, explicitly\footnote{The  result for $A_{41}(1^-2^-3^-4^+5^\varphi)$ is obtained by the exchange of the angle and square brackets.}
\begin{equation}
       A_{41}(1^+2^+3^+4^-5^\varphi) = \frac{4}{\Lambda^9}   s_{15} \langle 14 \rangle^2 [12]^2 [53]^2  +\text{perm}(1,2,3)\,.
\end{equation}
Note that the calculations of the 6pt amplitudes (at least of some of them) were necessary to reduce the system of parameters to one variable. In the next subsection we will try to understand this behavior from the another point of view based on recursion relations.

\subsection{Soft recursion relations}

From the previous subsection we have a strong hint that there is a unique scalar-vector theory with Galileon power-counting which obeys the soft theorems (\ref{eq:ansatzphotons}) and~(\ref{eq:ansatzop3}). We will postpone the actual construction of its Lagrangian based on a symmetry to the next section. Let us now assume its existence and discuss deeper the general properties of its amplitudes.
We will start with a proof that the amplitudes of this theory are fully reconstructible using the soft BCFW-like recursion.
Assume the amplitude
$A_{n_{\gamma },n_{\varphi }}(I^{+},I^{-},I^{\varphi })$ with $n_{\gamma }$ vector particles (photons) and $n_{\varphi}$ scalars. Provided  $n_{\gamma }+n_{\varphi }\geq 6$, we can apply the all-line soft shift \cite{Cheung:2015ota}
\begin{alignat}{2}
|\widehat{i}\rangle  &= (1-b_{i}z) |i\rangle \qquad &&\mathrm{for } \;\,i\in I^{+}  \notag \\
|\widehat{i}] &= ( 1-b_{i}z ) |i]&&\mathrm{for }\;\,i\in I^{-} 
\notag \\
\widehat{i} &= (1-a_{i}z)\, i&&\mathrm{for }\;\,i\in I^{\varphi}\,,
\end{alignat}
where $a_{i}$ and $b_{i}$ are (partially) fixed by the momentum conservation. Under this
shift, the amplitude scales for $z\rightarrow \infty $ as
\begin{equation}
\widehat{A}_{n_{\gamma }n_{\varphi }}\left( z\right) =O\left( z^{2\left(
n_{\gamma }+n_{\varphi }\right) -2-n_{\gamma }}\right) =O\left(
z^{2n_{\varphi }+n_{\gamma }-2}\right) 
\end{equation}
since, according to the power-counting $\rho =2$, the degree of homogeneity in momenta is $2\left( n_{\gamma
}+n_{\varphi }\right) -2$  and the
shift was chosen in such a way that the polarization of the photons are
not shifted\footnote{%
Note that the little group scaling requires undeformed $|i]|i]$ factor for
each positive helicity photon and similarly undeformed $|i\rangle |i\rangle $
for each negative helicity photon.}. The shift allows to probe the soft
limit of all the lines and including the corresponding soft factor, the
function
\begin{equation}
f_{n_{\gamma }n_{\varphi }}\left( z\right) =\frac{\widehat{A}_{n_{\gamma
}\,n_{\varphi }}\left( z\right) }{\prod\limits_{i=1}^{n_{\varphi }}\left(
1-a_{i}z\right) ^{3}\prod\limits_{j=1}^{n_{\gamma }}\left( 1-b_{j}z\right) }
\label{fnnz}
\end{equation}
scales for $z\rightarrow \infty$ as 
\begin{equation}
f_{n_{\gamma }n_{\varphi }}\left( z\right) =O\left( z^{-n_{\varphi
}-2}\right) .
\end{equation}
Therefore, there is no residue at infinity. The residue at $z=1/a_{i}$ vanishes due to the Special Galileon soft theorem, which ensures the scaling 
$\widehat{A}_{n_{\gamma },n_{\varphi }}(z) =O((1-a_{i}z)^{3})$. 
Although the amplitude $\widehat{A}_{n_{\gamma },n_{\varphi }}(z)$ does not vanish for $z=1/b_{i}$,
the corresponding residue is known in terms of the lower point amplitudes as
a consequence of the soft theorem (\ref{eq:ansatzphotons}). For instance for $i\in I^{+}$
\begin{equation}
\mathrm{res}\left( f_{n_{\gamma }n_{\varphi }} ,\frac{1}{%
 b_{i}}\right) =\frac{1}{ b_{i}\Lambda^3}\left[ \sum\limits_{j\in
I^{+}}\left[ i,j\right] ^{2}\widehat{A}_{n_{\gamma }-2,n_{\varphi
}+1}^{j^{+}\rightarrow j^{\varphi }}\left( \frac{1}{b_{i}}\right) -\sum\limits_{j\in I^{\varphi }}\left[ i,\widehat{j}\right] ^{2}%
\widehat{A}_{n_{\gamma },n_{\varphi }-1}^{j^{\varphi }\rightarrow j^{-}}\left( 
\frac{1}{b_{i}}\right) \right] 
\label{residue}
\end{equation}
and similarly for $i\in I^{-}$. As usual, the only remaining poles of $f_{n_{\gamma },n_{\varphi }}(z) $ are the unitarity poles of the
deformed amplitude and the corresponding residue are known from
factorization in terms of products of lower point amplitudes. We can
therefore apply the residue theorem on the meromorphic function $%
f_{n_{\gamma }n_{\varphi }}(z)/z$ and reconstruct the
amplitude $A_{n_{\gamma }n_{\varphi }}=\widehat{A}_{n_{\gamma }n_{\varphi
}}(0) $ recursively,
\begin{equation}
   \lim_{R\to\infty} \oint_{|z|=R} \frac{dz}{z} f_{n_{\gamma }n_{\varphi }}(z) = 0\quad \rightarrow \quad A_{n_{\gamma}n_{\varphi}} = - \sum_j \mathrm{res}\left( f_{n_{\gamma }n_{\varphi }} ,\frac{1}{b_{j}}\right) - \sum_{{\cal{F}}} \mathrm{res}\left( f_{n_{\gamma }n_{\varphi }}, z_{\cal{F}}\right).
   \label{Cauchy1}
\end{equation}
 The second term in the sum corresponds to standard factorization poles for $P_{\cal{F}}(z_{\cal{F}})^2=0$, where $P_{\cal{F}}^2$ is the propagator denominator in the factorization channel ${\cal F}$. The seed amplitudes for the recursion are given in~(\ref{eq:4ptinputs}) with~(\ref{eq:sol4pt}).

Let us point out that also the 5pt amplitudes can be fixed uniquely in terms of 4pt
ones. In this case, the all-line shift is not possible, but we can
always apply a three-line soft shift with two compensating momenta or a four-line soft shift with one compensating momentum. These correspond to the so-called all-but-two-line and all-but-one-line shifts, respectively (for details we refer to \cite{Cheung:2016drk}). 
Indeed, the shifted amplitude scales in both cases as $\widehat{A}_{n_{\gamma},5-n_{\gamma }}(z) =O(z^{8-n_\gamma})$ or better and
since $n_{\gamma }$ is always even, it is enough to use the three-line soft shift, i.e. more precisely $3\varphi$ soft shift for $A_{05}$ amplitude, $2\varphi +1\gamma $ or $3\varphi$ soft shift for $A_{23}$ amplitude
and $1\varphi +2\gamma $ soft shift for $A_{41}$. In the latter two cases
we have to choose carefully the type of the soft shift for various helicity
configurations. 

\subsection{Soft bonus relations}

In the previous subsection we have used the analytical properties of shifted amplitudes to reconstruct any higher-point amplitudes from the 4pt seeds. However, the role of the soft recursion relations is much subtle and consequently more important. This comes from the so-called {\emph{bonus relations}}  which serve as a consistency check on the inputs. Let us explain briefly where they come from. As already mentioned, the maximal soft shift we can employ on the 5pt amplitudes comes from the all-but-one-line shift, where 4 legs are simultaneously accessible for the soft limits. Note that all functions $f_{n_\gamma,n_\varphi}(z)$ given by (\ref{fnnz}) scale at least as $1/z^2$, and so for all 5p amplitudes $A_5(z)$ we don't need the $1/z$ factor in the Cauchy's integral~(\ref{Cauchy1}). 
If we drop the $1/z$ factor, we will not create a residue at infinity and thus we can apply the residue theorem directly to $f_{n_\gamma,n_\varphi}(z)$ and obtain relations among the 4pt vertices only. The similar analysis for the 6pt amplitudes leads to the same conclusion -- all amplitudes (including the 6-photons) have the large $z$ behavior at least as $1/z^2$. This can be easily generalized for higher points and we can summarize:
\begin{equation}
    \lim_{R\to\infty}\oint_{|z|=R} dz\,f_{n_\gamma n_\varphi}(z)  = 0\,,\qquad \text{for  } n_\gamma+n_\varphi\geq 5\,.
\end{equation}
Starting with $n_\gamma+n_\varphi=5$ we get that the sum of residua $f_{n_\gamma n_\varphi}$ includes only the lower-point amplitudes i.e. the 4pt vertices. If we use the general inputs as summarized in~(\ref{eq:4ptinputs}) it must give us relations among them. After this first iteration we found that it reduced the seven 4pt parameters down to three. Now, turning attention to the second iteration, i.e.  on the 6pt amplitudes, we got that all the 5pt vertices are fixed based on the 4pt inputs. So studying bonus relations with $n_\gamma+n_\varphi=6$ leads again to relations among the 4pt inputs. After performing this iteration we found out that only one parameter is left and the solution agrees with the direct calculation summarized in~(\ref{eq:sol4pt}).

We have thus used a second method which gave us exactly same result as a direct ``gluing'' of amplitudes. The advantage of the  method based on the bonus relations is mainly in the fact that we don't need to create an explicit basis for the higher point vertices. For the 5pt and 6pt this was still possible, however, for the higher ones this would be more and more difficult and, thus, the bonus relation method would be the only real option. In both cases we have come to a theory that has a chance to be an exceptional EFT, i.e. theory which is uniquely fixed by a single coupling constant.

\subsection{Generalization}

We have limited our discussion so far to a special form of the scalar soft limit~(\ref{eq:ansatzop3}). Now let us briefly focus on the general case of the soft scalar limit. We know that instead of $O(p^3)$ behavior in~(\ref{eq:ansatzop3}) we can demand general soft scalar limit with a non-zero right-hand side, given in~(\ref{general soft scalar theorem}). Now we may repeat the previous steps and try to reconstruct the theory again bottom up. There is, however, one technical difficulty. One can easily check that not all 5pt amplitudes are reconstructible anymore. To be more specific: the all-scalar 5pt amplitude cannot be reconstructed from the 4pt ones, but instead, it serves as one of the inputs for the 6pt (and higher-point) amplitudes.
We have already mentioned that there are two basis monomials for this amplitude (see ~(\ref{eq:5ptbaze})), but one can be eliminated by the scalar soft limit~(\ref{general soft scalar theorem}). For this combination, the right-hand side of the soft theorem is zero (trivially due to the momentum conservation) and hence this vertex must be equivalent to the 5pt Galileon amplitude
\begin{equation}
    A_{05}(1^\varphi 2^\varphi 3^\varphi 4^\varphi 5^\varphi) = 24 c_{05} \det \{s_{ij}\}_{i,j=1}^{4} = c_{05} s_{12}s_{13}s_{14}s_{15} + \text{perm(1,2,3,4,5)}\,.
\end{equation}
All other 5pt and higher-pt amplitudes are, however, reconstructible, i.e. they are zero, or they can be expressed by the seven 4pt constants, one 5pt constant and the  parameters of~the soft theorems  (\ref{general soft photon theorem}) and~(\ref{general soft scalar theorem}). There are only three such  parameters as the momentum conservation can be used to eliminate $c$ or $d$. We will take the following combination: 
\begin{equation}
    a,\;b\;\text{ and  }(d-c)
\end{equation}
as the independent choice.

Let us summarize what we have obtained after performing the same steps as for the special case, i.e. after testing all 5pt and 6pt amplitudes. There are two non-trivial cases:
\begin{align*}
    &\text{I)} &&a\neq -b,\; (d-c) = 0,   &&c_{22}''=0, &&c_{40}=c_{40}'= ab, &&c_{40}'' =0,\\
    &\text{II)} &&a=-b,\; (d-c)\neq 0, &&c_{22}''=\tfrac12 b(d-c),&&c_{40}=c_{40}'= -b^2,&&c_{40}'' =0\,.
\end{align*}
So in both cases the parametric space is reduced to six parameters: two soft-theorem constants ($a,b$ or $a=-b,d-c$), three 4pt constants $c_{04}$, $c_{22}$, $c_{22}'$ and one 5pt constant $c_{05}$. Note that these are only candidates for the solution, as we should check the self-consistency (stability of the solutions) by going to higher orders. In fact calculating all the 7pt amplitudes and one 8pt (the helicity-conserving amplitude $A_{80}$), we found out that the first solution trivializes and only the second  survives. We will see this is in agreement with results of the next section.

Before concluding this section let us mention one technical aspect which has never been discussed in literature. It concerns the above construction of the 5pt amplitudes based on the general soft scalar limit. As we have mentioned, not all 5pt amplitudes are reconstructible and it can be easily shown that those 5pt amplitudes that are reconstructible give no room for the bonus relations. Now, taking the amplitude $A_5(1^{+}2^{+}3^{+}4^{-}5^\varphi)$, we have found out that not-only this amplitude is completely fixed by the 4pt vertices (as it should be), but as a by-product we got one relation on the 4pt vertices themselves. How is this possible and how it can be explained using the general soft bootstrap reconstruction; does it mean that we have missed some possible shift for the 5pt kinematics, as summarized in \cite{Cheung:2016drk}? The answer is both, no and yes. It is true that it is not possible to shift simultaneously all five legs, as would be needed for the lacking bonus relation. However, the trick is, we can shift twice for the same kinematics, so effectively we {\it can\/} shift all five legs. First, we can shift say legs $1^{+},2^{+},3^{+},5^\varphi$ and then legs $1^{+},2^{+},4^{-},5^{\varphi}$. Both shifts must produce the same kinematics for $z=0$. In such a case we have two non-trivial and non-equivalent ways how to construct the amplitude $A_5$, schematically
\begin{equation}
    \text{soft bootstrap  } \Rightarrow \qquad A_5(0)|_{\text{shift }1} = A_5(0)|_{\text{shift }2}
\end{equation}
and this leads to the desired constrain(s) on the inputs, i.e. on the 4pt vertices.

To summarize the soft bootstrap procedure produced unique amplitudes for massless vectors and scalars with $\rho=2$ power-counting that satisfy the soft theorems (\ref{rho2 soft theorem}). This is a smoking gun for the existence of a particular theory with some underlying symmetry. On top of that, the situation is similar to the pure scalar Galileon, where for the general theory there exists a special choice of couplings leading to the Special Galileon, which possesses a hidden symmetry. This seems to be also true in the scalar-vector sector as hinted by the soft bootstrap results.  We will elaborate on this in the next section.

\section{The scalar-vector Galileon theory}

Based on the results of the last section, we are now looking for a coupled scalar-vector theory with $\rho=2$ power-counting, with scattering amplitudes satisfying the soft theorems (\ref{eq:ansatzphotons}) and (\ref{eq:ansatzop3}). 

\subsection{The coupling of Special Galileon to massless vector}

Our starting point in the identification of this theory is the purely scalar sector. It must be given by the Special Galileon theory in $D=4$:
\begin{equation}
\mathcal{L}_{sGal}=-\frac{1}{2}\varphi \square \varphi \pm\frac{1}{24\alpha
^{2}}\varphi \left( \left( \square \varphi \right) ^{3}
+2 \partial_\mu\partial^\nu\varphi \partial_\nu\partial^\rho \varphi\partial_\rho\partial^\mu\varphi 
-3 \square\varphi \partial_\mu\partial_\nu\varphi \partial^\mu\partial^\nu\varphi \right)\,,
\label{sGal Lagrangian}
\end{equation}
which leads to the desired soft limit behavior~(\ref{eq:ansatzop3}).
Here $\alpha$ is a dimensionful parameter with $[\alpha]=3$ which sets the scale of non-linearity and plays the role of a coupling constant. The two possible signs correspond to two branches of the Special
Galileon Lagrangians, which can be connected by means of analytic
continuation $\alpha \rightarrow \mathrm{i}\alpha$. According to the general theorem proved in \cite{Cheung:2016drk}, the soft behavior of the amplitudes can be understood as a consequence of some generalized
polynomial shift symmetry. In the case of Special Galileon Lagrangian (\ref{sGal Lagrangian}), such a symmetry can be
treated as a simultaneous transformation of the coordinates and the Galileon
field \cite{Hinterbichler:2015pqa,Novotny:2016jkh}
\begin{equation}
x^{\prime \mu } =x^{\mu }+G^{\mu \nu }\partial _{\nu }\varphi \left(
x\right),   \qquad 
\varphi ^{\prime }\left( x^{\prime }\right)  =\varphi (x) -\frac{1}{2}G^{\alpha \beta }\left( \pm \alpha ^{2}x_{\alpha }x_{\beta
}-\partial _{\alpha }\varphi \partial _{\beta }\varphi \right) 
\label{sGal transformation}
\end{equation}
with an infinitesimal parameter $G_{\mu \nu }$ with properties 
$G^{\mu\nu }=G^{\nu \mu }$, 
$\eta _{\mu \nu }G^{\mu \nu }=0$. The $\pm$ signs correspond to two branches discussed above.

Our task is to construct the most general Lagrangian including the
fields $\varphi(x)$ and $A_\mu(x)$ which is
invariant with respect to the transformation (\ref{sGal transformation}) enlarged appropriately to
the field $A_\mu(x)$. The geometrical interpretation of the
Special Galileon developed in \cite{Novotny:2016jkh} proved to be particularly useful for our purposes. Within this approach, the Special Galileon field is treated as the only nontrivial degree of freedom corresponding to a particular gauge description of a configuration of a $D-$dimensional Lorentzian brane in $2D$ dimensional flat target space with
pseudo-Riemannian metric with a signature either $\left(2,2D-2\right) $ or $\left( D,D\right)$ according to the branch. The
Special Galileon symmetry can be then interpreted as a linear isometry of
the target space and its non-linear nature when acting on the Special
Galileon field is a consequence of a particular gauge fixing. Such
picture allows us to easily construct the covariant building blocks with
respect to this symmetry. Namely, the basic objects are the induced metric on the brane
\begin{equation}
g_{\mu \nu }=\eta _{\mu \nu }\pm\frac{1}{\alpha ^{2}}\partial _{\mu }\partial
\varphi \cdot \partial \partial _{\nu }\varphi ,
\end{equation}
which transforms as a rank two tensor under (\ref{sGal transformation}), the inverse metric $g^{\mu\nu}$, the corresponding covariant derivative with the Christoffel symbol $\Gamma_{\rho \mu \nu }$, and the extrinsic curvature $\mathcal{K}_{\mu \nu \alpha }$,
\begin{equation}
\Gamma _{\rho \mu \nu }=\pm\frac{1}{\alpha ^{2}}\partial _{\mu }\partial _{\nu
}\partial \phi \cdot \partial \partial _{\rho }\phi ,
\qquad \mathcal{K}_{\mu \nu \alpha }=-\frac{1}{\alpha }\partial _{\alpha }\partial
_{\mu }\partial _{\nu }\phi \,,
\end{equation}
where $\mathcal{K}_{\mu \nu \alpha }$ transforms under (\ref{sGal transformation}) as a rank three tensor. Next building bock is the scalar $\sigma$ which is defined differently for two branches mentioned above, namely
\begin{equation}
\sigma=  \frac{\alpha }{2\mathrm{i}}\ln \frac{\det \left( \eta +\frac{\mathrm{i}}{%
\alpha }\partial \partial \varphi \right) }{\det \left( \eta -\frac{\mathrm{i
}}{\alpha }\partial \partial \varphi \right) }\qquad \mbox{and} \qquad
 \sigma=   \frac{\alpha }{2}\ln \frac{\det \left( \eta +\frac{1}{\alpha }\partial
\partial \varphi \right) }{\det \left( \eta -\frac{1}{\alpha }\partial
\partial \varphi \right) },
\end{equation}
for the ``plus'' and ``minus'' branches, and it transforms as an invariant with respect to the Special Galileon symmetry
\begin{equation}
    \sigma^{\prime}(x^{\prime})=\sigma(x).
\end{equation}
The last building block is the invariant measure $\mathrm{d}^{D}x\sqrt{|g|}$ on the brane, where $g=\det g_{\mu \nu }$. 
Then any diffeomorphism invariant built from the above building blocks is
automatically invariant with respect to the Special Galileon symmetry (\ref{sGal transformation}). This suggests to enlarge the Special Galileon
symmetry to the vector field $A_{\mu }$ according to
\begin{equation}
A_{\mu }^{\prime }\left( x^{\prime }\right) =\frac{\partial x^{\alpha }}{%
\partial x^{\prime \mu }}A_{\alpha }\left( x\right) =A_\mu(x)
-G^{\alpha \beta }\partial _{\beta }\partial _{\mu }\varphi (x)
A_{\alpha }\left( x\right) .
\label{A transformation}
\end{equation}
Scalar-vector Lagrangians invariant with respect to (\ref{sGal transformation}) and (\ref{A transformation}) lead then automatically to the amplitudes satisfying the enhanced $O(p^3)$ soft scalar limit.

From the above building blocks we can  construct the following minimal invariant Lagrangian which corresponds
to the coupling of the vector field $A_\mu$ to the Special Galileon
\begin{equation}
\mathcal{L}_{V }=
\mathcal{L}_{sGal}
-\frac{1}{4}\sqrt{\left\vert g\right\vert }V\left( \frac{%
\sigma }{\alpha }\right) F_{\mu \alpha }F_{\nu \beta }g^{\mu \nu }g^{\alpha
\beta } \,,\label{vecskal}
\end{equation}
where $V$ is an arbitrary potential given by the expansion
\begin{equation}
V\left( \frac{\sigma }{\alpha }\right) =1+\sum_{n=1}^{\infty }\frac{1}{n!}%
v_{n}\left( \frac{\sigma }{\alpha }\right) ^{n}.
\end{equation}
The canonical normalization of the vector kinetic term requires $V\left( 0\right) =1$ and hermiticity needs $v_{n}$ to be real.
Note also that the Lagrangian has the same power-counting as the Galileon, i.e. each vertex has $\rho=2$.
The minimal form of such a Lagrangian with $V=1$ has been studied in \cite{Bonifacio:2019rpv}.

Of course, ${\cal{L}}_V$ is not the most general 
invariant Lagrangian  with the Galileon power-counting since we can freely add non-minimal invariants
which are higher order in $F_{\mu\nu}$ of the  schematic form 
\begin{equation}
{\cal L} =  V(\sigma) \cdot \cder^{n_{\cder}}\cdot \ctns^{n_\ctns}\cdot F^{n_F} \qquad\mbox{with}\qquad n_{F}-2=n_{\cder}+n_{\ctns}\,,
\end{equation}
where $\cder_\mu$ is the covariant derivative associated with the metric $g_{\mu\nu}$. The constraint on powers $n_{F}$, $n_{\cder}$, $n_{\ctns}$ is necessary to preserve $\rho=2$ power-counting. The minimal case (\ref{vecskal}) has $n_F=2$, $n_{\cder}=n_{\ctns}=0$. The  terms quartic in $F_{\mu\nu}$ are of the schematic form $V F^4\ctns^2$ (12 independent terms), $V (\cder\ctns) F^4$ (4 terms), $V (\cder F)^2 F^2$ (9 terms), $V (\cder F) F^3 \ctns$ (10 terms), and $V (\cder^2 F) F^3$ (5 terms), and so on.

Note that {\emph{any}} scalar-vector theory invariant with respect to the above Special Galileon
symmetry (\ref{sGal transformation}), (\ref{A transformation}) has to include the minimal Lagrangian  $\mathcal{L}_{V }$ with some choice of $V$ in order to get the 
kinetic term for the vector field $A_{\mu}$.
Even in this minimal case  we are left with an infinite number of independent parameters $v_n$.
In order to get a unique and on-shell reconstructible theory based on $\mathcal{L}_{V }$, we need therefore to fix the form of the potential $V$.
This confirms the above results based on amplitude bootstrap: the scalar-vector theory cannot be fixed uniquely using only the $O(p^3)$ single scalar soft limit. 

\subsection{The Special scalar-vector Galileon}

Assuming  the minimal Lagrangian $\mathcal{L}_{V }$ and fixing the  ``minus'' branch of the
Special Galileon, the choice of the potential $V(x)$ according to 
\begin{equation}
V\left( x\right) =\mathrm{e}^{x}
\end{equation}
is very special. In this case, the action of the theory has additional symmetry which together with the Special Galileon symmetry fixes the tree-level $S-$matrix uniquely, and it corresponds to the solution we found using the bootstrap method.

To reveal this additional symmetry we can use the Galileon duality  transformation (see \cite{deRham:2013hsa} and \cite{Kampf:2014rka} for details) extended appropriately to the vector filed $A_{\mu}$, which relates  the original theory to an equivalent dual theory with the same $S-$matrix. For a particular choice of the duality transformation the additional symmetry will be manifest in the dual formulation of the theory.
This duality transformation can be described as follows. Let us rename the original fields and coordinates in the Lagrangian (\ref{vecskal}) as
\begin{equation}
x\to y,\,\,\,\varphi(x)\to \rho(y),\,\,\,A_{\mu}(x)\to B_{\mu}(y),\,\,\,F_{\mu\nu}(x)\to G _{\mu\nu}(y),  
\end{equation}
then the new fields and coordinates, which we denote again as $x$, $\varphi(x)$ and $A_{\mu}(x)$, are given in terms of the original ones as
\begin{equation}
x =y-\frac{1}{\alpha }\partial \rho(y),\,\,\,\,
\varphi(x)  =\rho (y) -
\frac{1}{2\alpha }\partial \rho (y) \cdot \partial \rho
\left( y\right),\,\,\,\,
A_{\mu  }\left( y\right)  =\frac{\partial y^{\alpha
}}{\partial x^{ \mu }}B_{\alpha }\left( y\right). 
\end{equation}
This transformation leads to the dual action $S^{\prime}[\varphi,F_{\mu\nu}]$ defined as $S^{\prime}[\varphi,F_{\mu\nu}]=S[\rho,G_{\mu\nu}]$. Explicitly we get
\begin{equation}
S^{\prime }[\varphi,F_{\mu\nu}]=
\int \mathrm{d}^{4}x\mathcal{L}^{\prime}_{sGal}-\frac{1}{4}\int \mathrm{d}^{4}x\sqrt{\left\vert
g^{\prime }\left( x\right) \right\vert }\,\mathrm{exp}\left(\frac{\sigma
^{\prime }\left(x\right) }{\alpha }\right)F_{\mu \alpha }\left( x\right) F_{\nu \beta }\left( x\right)
g^{\prime \mu \nu }\left( x\right) g^{\prime \alpha \beta }\left(
x\right). 
\end{equation}
Since the original scalar-vector interaction Lagrangian was build from geometrical objects which transform covariantly under the change of the coordinates, the general form of this part of the Lagrangian is preserved under the duality transformation, while the Special Galileon Lagrangian is replaced by the dual Special Galileon Lagrangian
 $\mathcal{L}^{\prime}_{sGal}$ \cite{Hinterbichler:2015pqa,Roest:2020vny}
\begin{eqnarray}
\mathcal{L}_{sGal}^{\prime } &=&\sum\limits_{n=1}^{4} 
 \frac{1}{(n+1)!(4-n)!}\left(\frac{2}{\alpha }\right) ^{n-1}\varphi (x)\,\varepsilon ^{\mu _{1}\mu _{2}\mu _{3}\mu _{4}}\varepsilon ^{\nu _{1}\nu
_{2}\nu _{3}\nu _{4}}\prod\limits_{i=1}^{n}\partial _{\mu _{i}}\partial_{\nu _{i}}\varphi(x)\prod\limits_{j=n+1}^{4}\eta _{\mu _{j}\nu _{j}}.\notag\\
\end{eqnarray}
The dual transformed metric $g^{\prime}_{\mu\nu}$ (in the above formula $g^{\prime\mu\nu}$ is the corresponding inverse) and scalar invariant $\sigma^{\prime}$  can be expressed in terms of the dual field $\varphi$ as
\begin{equation}
g_{\mu \nu }^{\prime }\left( x \right)  =\eta _{\mu \nu }+\frac{2}{\alpha }\partial _{\mu }\partial_{\nu }\varphi
\left( x\right),\,\,\,\,  
\sigma^{\prime }\left(x\right)  =\frac{\alpha }{2}\left\vert
g^{\prime }\left( x\right)\right\vert.
\end{equation}
After some manipulation\footnote{Here we use the formula $\varepsilon^{\mu_1\mu_2\alpha
_1\alpha_2}\varepsilon^{\nu_1\nu_2\beta_1\beta_2}g^{\prime}_{\alpha_1\beta_1}g^{\prime}_{\alpha_2\beta_2}=-2| g^{\prime}|(g^{\prime\mu_1\nu_1}g^{\prime\mu_2\nu_2}-g^{\prime\mu_1\nu_2}g^{\prime\mu_2\nu_1})$.} we find for the dual scalar-vector interaction Lagrangian
\begin{eqnarray}
{\cal{L}}_{int}^{\prime}&=&-\frac{1}{4}\sqrt{\left\vert
g^{\prime } \right\vert }\,\mathrm{exp}\left({\frac{\sigma
^{\prime } }{\alpha }}\right)F_{\mu \alpha } F_{\nu \beta }
g^{\prime \mu \nu } g^{\prime \alpha \beta }=\frac{1}{4}\widetilde{F}^{
\mu \nu } \widetilde{F}^{ \alpha \beta
} g_{\mu \alpha }^{\prime } g_{\nu \beta }^{\prime }  \,,
\end{eqnarray}
where $\widetilde{F}^{\mu\nu}=\frac{1}{2}\varepsilon^{\mu\nu\alpha\beta}F_{\alpha\beta}$. Explicitly we get
\begin{equation}
{\cal{L}}_{int}^{\prime}    = -\frac{1}{4} F_{\mu
\nu }F^{ \mu \nu }+\frac{1}{4\alpha }\varepsilon ^{\alpha
\beta \mu _{1}\mu _{2}}\varepsilon ^{\gamma \delta \nu _{1}\nu
_{2}}F_{\alpha \beta }^{ }F_{\gamma \delta }\left( \partial
_{\mu _{1}}\partial _{\nu _{1}}\varphi\eta _{\mu _{2}\nu _{2}}+%
\frac{1}{\alpha }\partial _{\mu _{1}}\partial _{\nu _{1}}\varphi \partial _{\mu _{2}}\partial _{\nu _{2}}\varphi \right).
\label{Lgammagamma}
\end{equation}
Note that the last form of the Lagrangian ${\cal{L}}_{int}^{\prime}$ 
can be identified with a special case of the shift symmetric scalar-vector Lagrangians discussed in \cite{Bonifacio:2019rpv} which are related to the scalar-vector sector of the flat space decoupling limit of the massive spin-2 pseudo-linear theory. It appears also as a decoupling limit of the special case of the generalized Proca Lagrangians introduced in \cite{Heisenberg:2017mzp}.

Written in the form (\ref{Lgammagamma}), ${\cal{L}}_{int}^{\prime}$ is manifestly invariant (up to the total derivative) with respect to the shift
\begin{equation}
\delta F_{\alpha \beta }=m_{\alpha \beta }\,,
\label{F shift}
\end{equation}
where $m_{\alpha \beta }$ is the infinitesimal antisymmetric tensor. 
Such a shift is induced by the following linear shift of the four-potential
\begin{equation}
\delta A_{\mu }=\frac{1}{2}x^{\alpha }m_{\alpha \mu }.
\label{A shift}
\end{equation}
The original Special Galileon symmetry (\ref{sGal transformation}) can be expressed in the dual formulation as
\begin{eqnarray}
x^{\prime\mu}= x^{ \mu } -G^{\mu \nu }x_{\nu},\,\,\,\,
\varphi^{\prime}(x^{\prime})=
\varphi(x)+\frac{\alpha}{2}G^{\alpha \beta }x_{\alpha} x_{\beta} 
\label{dual sGal transformation}
\end{eqnarray}
and its extension (\ref{A transformation}) to the vector field $A_{\mu}$ reads now
\begin{equation}
A_{\mu}^{\prime}(x^{\prime})=A_{\mu}(x)+G^{\alpha}_{\,\,\,\mu}A_{\alpha}(x)\,.
\end{equation}
This transformation remains to be a symmetry of the dual action $S^{\prime}[\varphi,F_{\mu\nu}]$.
Since the $S-$matrix in both the original and dual formulations are the same, the amplitudes are unchanged and obey the enhanced $O(p^3)$ Adler zero for soft scalars. The shift symmetry (\ref{F shift}), (\ref{A shift}) results in the  additional soft-vector theorems which is a special case of the general one (\ref{general soft photon theorem}) (for the  proof see Appendix \ref{soft_theorems}). For a helicity plus soft vector particle we get
\begin{eqnarray}
\lim_{p\rightarrow 0}A_{n+1}\left( p^{+},I^{+},I^{-},I^{\varphi }\right) &=&%
-\frac{1}{2\alpha }\sum\limits_{i\in I^{+}}\left[ p,i\right] ^{2}A_{n}\left(
I^{+},I^{-},I^{\varphi }\right) _{i^{+}\rightarrow i^{\varphi }}\notag\\
&&+\frac{1}{%
2\alpha }\sum\limits_{i\in I^{\varphi }}\left[ p,i\right] ^{2}A_{n}\left(
I^{+},I^{-},I^{\varphi }\right) _{i^{\varphi }\rightarrow i^{-}},
\label{soft photon theorem}
\end{eqnarray}
and similarly for the helicity minus soft vector particle. We checked explicitly that for $\alpha=\Lambda^3/2$ the Lagrangian (\ref{Lgammagamma}) generates the same amplitudes we found using the bootstrap method and it is indeed identified with the unique solution to the soft limit constraints (\ref{eq:ansatzphotons}) and (\ref{eq:ansatzop3}) we found earlier.

\subsection{General scalar-vector Galileon}\label{sec:genscalgal}

The Lagrangian (\ref{Lgammagamma}) is not the most general form which emerges as the decoupling limit of the massive spin-2 pseudolinear theory or generalized Proca theory. Similarly, the form of the soft theorems satisfied by the corresponding amplitudes are not of the most general form (\ref{general soft photon theorem}) and (\ref{general soft scalar theorem}) discussed in Section \ref{soft theorems}.
In this subsection, we will discuss other examples of theories with more general soft theorems. 

Assume the Lagrangian which couple the general Galileon with massless vector that has been derived in \cite{Bonifacio:2019hrj} as the most general scalar-vector sector of the decoupling limit of the massive spin-2 pseudolinear theory, namely
\begin{eqnarray}
\mathcal{L} &=&\frac{1}{2}\partial \varphi \cdot \partial \varphi
+\sum\limits_{n=2}^{4}\frac{d_{n+1}}{\Lambda^{3(n-1)}}\,\varphi \varepsilon ^{\mu _{1}\mu _{2}\mu
_{3}\mu _{4}}\varepsilon ^{\nu _{1}\nu _{2}\nu _{3}\nu
_{4}}\prod\limits_{i=1}^{n}\partial _{\mu _{i}}\partial _{\nu _{i}}\varphi
\prod\limits_{j=n+1}^{4}\eta _{\mu _{j}\nu _{j}}\label{general Galileon Lagrangian}\\
&&-\frac{1}{4}F_{\mu \nu }F^{ \mu \nu }+\frac{1}{4}\varepsilon ^{\alpha
\beta \mu _{1}\mu _{2}}\varepsilon ^{\gamma \delta \nu _{1}\nu
_{2}}F_{\alpha \beta }F_{\gamma \delta }\left( \frac{a_3}{\Lambda ^{3}}%
\partial _{\mu _{1}}\partial _{\nu _{1}}\varphi \eta _{\mu _{2}\nu
_{2}}+\frac{a_4}{\Lambda ^{6}}\partial _{\mu _{1}}\partial _{\nu _{1}}\varphi
\partial _{\mu _{2}}\partial _{\nu _{2}}\varphi \right). \notag
\end{eqnarray}
It is manifestly invariant (up to the total derivative) with respect
to the shift 
\begin{equation}
\delta F_{\alpha \beta }=m_{\alpha \beta } 
\end{equation}
 and with respect
to the Galileon symmetry%
\begin{equation}
\delta \varphi =a+b\cdot x.
\label{general galilleon symmetry}
\end{equation}
The first symmetry is responsible for the soft photon theorem (see Appendix \ref{soft_theorems} for more details)
\begin{eqnarray}
\lim_{p\rightarrow 0}A_{n+1}\left( p^{+},I^{+},I^{-},I^{\varphi }\right) &=&-%
\frac{a_3}{2\Lambda ^{3}}\sum\limits_{i\in I^{+}}\left[ p,i\right]
^{2}A_{n}\left( I^{+},I^{-},I^{\varphi }\right) _{i^{+}\rightarrow
i^{\varphi }}\notag\\
&&+\frac{a_3}{2\Lambda ^{3}}\sum\limits_{i\in I^{\varphi }}\left[
p,i\right] ^{2}A_{n}\left( I^{+},I^{-},I^{\varphi }\right) _{i^{\varphi
}\rightarrow i^{-}}\,.
\label{general soft vector theorem}
\end{eqnarray}
The second symmetry leads to a nontrivial soft scalar theorem (note that unlike the Special Galileon case we have uncorrelated
cubic vertices\footnote{In the case of the Special Galileon, $6\frac{d_3}{\Lambda^3}=a_3/\Lambda^3=1/2\alpha$ and the theorem is trivial due to the momentum conservation.}) of the form
\begin{equation}
A_{n+1}\left( p^{\varphi },I^{+},I^{-},I^{\varphi }\right) \overset{%
p\rightarrow 0}{=}\biggl[ -6\frac{d_{3}}{\Lambda^3}\sum\limits_{i\in I^{\varphi }}(p\cdot i) -\frac{a_3}{\Lambda ^{3}}\sum\limits_{i\in I^{+}\cup
I^{-}}( p\cdot i) \biggr] A_{n}\left( I^{+},I^{-},I^{\varphi
}\right) +O( p^{2}), 
\end{equation}
(for the proof see Section~\ref{ap:secsst}). The amplitudes of the theory with $n_{\varphi}+n_{\gamma}\geq 6$ are  fully reconstructible using the soft recursion relations. Indeed, using the all-line shift based on the above soft theorems, we get for $z\to\infty$
\begin{equation*}
f_{n_\gamma n\varphi}(z)=\frac{A\left( z\right) }{\prod\limits_{i=1}^{n_{\varphi }}\left(
1-a_{i}z\right) ^{2}\prod\limits_{j=1}^{n_{\gamma }}\left( 1-b_{j}z\right) }%
=O\left( \frac{z^{2(n_\varphi+n_\gamma)-2-n_{\gamma }}}{z^{2n_{\varphi }}z^{n_{\gamma }}}%
\right) =O\left( z^{-2}\right). 
\end{equation*}
Unlike the Special Galileon case, not all 5pt amplitudes are reconstructible, since the all-line soft shift is not available. 
However, we can use the soft four-line shift, which, as discussed in \cite{Cheung:2016drk} or in previous section, is available for $n_\gamma+n_\varphi\geq 5$.
Using the $1\gamma+3\varphi$ soft shift for $A_{23}$ and $3\gamma+1\varphi$ soft shift for $A_{41}$ we can reconstruct these amplitudes form the 4pt seeds.

We are left with just seven\footnote{Here we take into account various helicity configurations, however, we do not take the helicity flipped amplitudes as independent due to the parity conservation.} seed amplitudes of the schematic type $A_{04}$, $A_{22}$, $A_{40}$, and $A_{05}$. They are not all independent since they are expressed in terms of five independent couplings $d_3$, $d_4$, $d_5$, $a_3$ and $a_4$ only.

There is another interesting class of theories, which are related to the Lagrangian (\ref{general Galileon Lagrangian}). 
These theories are invariant with respect to one parametric generalization of the shift symmetry (\ref{A shift}) with the deformation parameter $\theta$,
\begin{equation}
\delta A_\mu=\frac{1}{2}m_{\alpha\mu}(x^\alpha+2\frac{\theta}{\Lambda^3}\partial^\alpha\varphi)\equiv \delta^{(0)}A_\mu+\theta\delta^{(1)}A_\mu,
\label{deformed shift}
\end{equation}
as well as with respect to the Galileon symmetry (\ref{general galilleon symmetry}). Such theories correspond to the scalar-vector sector of the decoupling limit of the massive gravity as discussed in details in \cite{Bonifacio:2019hrj}. The Lagrangian of these theories can be obtained from the ``seed'' Lagrangian (\ref{general Galileon Lagrangian}) by the deformation procedure which produces infinite tower of descendant terms from each scalar-vector interacting terms of the Lagrangian (\ref{general Galileon Lagrangian}).
Schematically,
\begin{equation}
 {\cal{L}}= {\cal{L}}_{Gal}+ \sum_{n=2}^{4}\sum_{k=n}^{\infty}  {\cal{L}}^{(k)}_n,
 \label{most general case}
\end{equation}
where the seeds ${\cal{L}}^{(n)}_n$ for $n=2,3,4$ are the vector kinetic term, the $2\gamma\varphi$ cubic term and $2\gamma2\varphi$ quartic interaction terms of the original Lagrangian  (\ref{general Galileon Lagrangian}), while ${\cal{L}}^{(k)}_n $ are their descendants (here $k$ denotes the  number of fields while $n$ denotes valence of the seed).
These satisfy the fundamental relations
\begin{equation}
 \delta^{(0)} {\cal{L}}^{(k)}_n=  \theta\delta^{(1)} {\cal{L}}^{(k-1)}_n ,
\end{equation}
which ensure the invariance with respect to (\ref{deformed shift}).
The solution to the above relations was found in \cite{Bonifacio:2019hrj}, where the reader can also find the explicit formulas which we do not reproduce here. Let us only mention the explicit form of the only new cubic vertex ${\cal{L}}^{(3)}_2$
 \begin{equation}
{\cal{L}}^{(3)}_2=\frac{\theta}{2\Lambda^3}F^{\mu\alpha}F^{\nu}_{\,\,\,\alpha}\partial_\mu\partial_\nu\varphi,
 \end{equation}
which is crucial for the derivation of the soft scalar theorem.
The resulting Lagrangian contains six free parameters, namely the Galileon couplings $d_3,d_4,d_5$, the seed couplings $a_3,a_4$ and the deformation parameter $\theta$. For this theory it can be proved the following modification of the soft theorems\footnote{The general proof will be published elsewhere.} (\ref{general soft vector theorem}), namely for helicity plus soft vector
\begin{eqnarray}
\lim_{p\rightarrow 0}A_{n+1}\left( p^{+},I^{+},I^{-},I^{\varphi }\right) &=&-%
\frac{2a_3-\theta}{4\Lambda ^{3}}\sum\limits_{i\in I^{+}}\left[ p,i\right]
^{2}A_{n}\left( I^{+},I^{-},I^{\varphi }\right) _{i^{+}\rightarrow
i^{\varphi }}\notag\\
&&+\frac{2a_3-\theta}{4\Lambda ^{3}}\sum\limits_{i\in I^{\varphi }}\left[
p,i\right] ^{2}A_{n}\left( I^{+},I^{-},I^{\varphi }\right) _{i^{\varphi
}\rightarrow i^{-}}
\label{general soft vector theorem deformed}
\end{eqnarray}
and similar one for helicity minus case, and for soft scalar
\begin{equation}
A_{n+1}(p^{\varphi },I^{+},I^{-},I^{\varphi }) \overset{p\rightarrow 0}{=}\Bigl[ -6\frac{d_{3}}{\Lambda^3}\sum\limits_{i\in I^{\varphi }}
(p\cdot i) -\frac{2a_3+\theta}{2\Lambda^{3}}
\sum\limits_{i\in I^{+}\cup I^{-}}(p\cdot i) \Bigr] A_{n}(I^{+},I^{-},I^{\varphi}) +O(p^{2}).
\label{general soft scalar deformed}
\end{equation}
This theory is a realization of the general soft theorems (\ref{general soft photon theorem}) and (\ref{general soft scalar theorem}).
As above, the amplitudes are reconstructible from the six parametric set of seed amplitudes $A_{04}$, $A_{22}$, $A_{40}$ and $A_{05}$.

\section{Summary of Galileon theories}\label{sec:sum}

The most general case of the scalar-vector Galileon theory, which is described by the Lagrangian (\ref{most general case}), can be uniquely defined as a theory satisfying the following generalized soft theorems, namely for the soft helicity plus vector
\begin{eqnarray}
\lim_{p\rightarrow 0}A_{n+1}\left( p^{+},I^{+},I^{-},I^{\varphi }\right) &=&
-\xi\sum\limits_{i\in I^{+}}\left[ p,i\right]
^{2}A_{n}\left( I^{+},I^{-},I^{\varphi }\right) _{i^{+}\rightarrow
i^{\varphi }}\notag\\
&&+\xi\sum\limits_{i\in I^{\varphi }}\left[
p,i\right] ^{2}A_{n}\left( I^{+},I^{-},I^{\varphi }\right) _{i^{\varphi
}\rightarrow i^{-}},
\label{eq:softx}
\end{eqnarray}
for the helicity minus soft vector
\begin{eqnarray}
\lim_{p\rightarrow 0}A_{n+1}\left( p^{-},I^{+},I^{-},I^{\varphi }\right) &=&
-\xi\sum\limits_{i\in I^{+}}\langle p,i\rangle
^{2}A_{n}\left( I^{+},I^{-},I^{\varphi }\right) _{i^{-}\rightarrow
i^{\varphi }}\notag\\
&&+\xi\sum\limits_{i\in I^{\varphi }}\langle p,i\rangle^{2}A_{n}\left( I^{+},I^{-},I^{\varphi }\right) _{i^{\varphi
}\rightarrow i^{+}}
\label{eq:softx1}
\end{eqnarray}
and for the soft scalar
\begin{equation}
A_{n+1}\left( p^{\varphi },I^{+},I^{-},I^{\varphi }\right) \overset{%
p\rightarrow 0}{=} \eta\sum\limits_{i\in I^{\varphi }}\left(
p\cdot i\right) A_{n}\left( I^{+},I^{-},I^{\varphi
}\right) +O\left( p^{2}\right)\,.
\label{eq:softy}
\end{equation}
The seeds for the soft recursion based on the above soft theorems are the 4pt amplitudes
\begin{equation}
\begin{split}
A_{04}(1^\varphi2^\varphi3^\varphi4^\varphi) &= c_{04} \, s_{12}s_{13}s_{23}\\
A_{22}(1^+2^-3^\varphi4^\varphi) &= c_{22}\, s_{34}[1|3|2\ra[1|4|2\ra\\
A_{22}(1^+2^+3^\varphi4^\varphi) &= c_{22}'\, (s_{13} s_{23}+s_{14} s_{24})[12]^2 + \frac{\xi\eta}{2}\, s_{12}^2 [12]^2 \\
A_{40}(1^+2^+3^-4^-)&= -\xi^2\, s_{12}[12]^2\la34\ra^2\\
A_{40}(1^+2^+3^+4^+)&= -\xi^2\, \bigl(s_{12}[12]^2[34]^2 + s_{13}[13]^2[24]^2 + s_{23} [23]^2[14]^2\bigr)\\
A_{40}(1^+2^+3^+4^-)&= 0
\end{split}
\end{equation}
and one 5pt amplitude 
\begin{equation}
    A_{05}(1^\varphi 2^\varphi 3^\varphi 4^\varphi 5^\varphi) = c_{05} s_{12}s_{13}s_{14}s_{15} + \text{perm(1,2,3,4,5)}\,.
\end{equation}
Altogether we have six free parameters: $\xi$, $\eta$, $c_{04}$, $c_{22}$,  $c_{22}'$ and $c_{05}$, which determine the theory uniquely. 
Their explicit form in terms of the Lagrangian parameters $a_3$, $a_4$, $d_3$, $d_4$, $d_5$ and $\theta$ can be found in Appendix~\ref{app:amp45}.
It is the most economic way how to represent this theory.  
Note that within the Lagrangian description we need to specify the couplings of an infinite tower of vertices.

A special choice of the parameters leads to an exceptional EFT, the {\emph{Special scalar vector Galileon}}.
This theory corresponds to
the fixing of the parameters in terms of just one free coupling $\alpha$ according to
\begin{equation}
    \xi = \frac{1}{2\alpha}\,,\;\eta=0 \qquad \text{and }\quad
c_{04} = c_{22} = \frac{1}{4\alpha^2}\,,\;
c_{22}'=c_{05}= 0  \,.
\end{equation}
In this case, the soft scalar theorem is even enhanced in comparison with~\eqref{eq:softy} with $\eta=0$, namely
\begin{equation}
A_{n+1}\left( p^{\varphi },I^{+},I^{-},I^{\varphi }\right) \overset{%
p\rightarrow 0}{=} O\left( p^{3}\right)\,.
\end{equation}
This theory can be expressed by a Lagrangian in a closed form. We have found two equivalent formulation. In the first form it is still expressed by infinitely many terms
\begin{align}
\mathcal{L}_{spec}&=-\frac{1}{2}\partial \varphi \cdot \partial \varphi-\frac{1}{24\alpha
^{2}}\varphi \left( \left( \square \varphi \right) ^{3}
+2 \partial_\mu\partial^\nu\varphi \partial_\nu\partial^\rho \varphi\partial_\rho\partial^\mu\varphi 
-3 \square\varphi \partial_\mu\partial_\nu\varphi \partial^\mu\partial^\nu\varphi \right)\notag\\
&-\frac{1}{4}\det \bigl(\eta +\frac{1}{\alpha }\partial\partial \varphi \bigr)
F_{\mu \alpha }F_{\nu \beta }
g^{\mu\nu} g^{\alpha\beta}\,,
\end{align}
as the inverse of metric is given by the following series
\begin{equation}
    g^{\mu\nu} = \eta^{\mu\nu} + \sum_{n=1}^\infty \frac{1}{\alpha^{2n}} [(\partial\partial\phi \cdot \partial \partial \phi)^n]^{\mu\nu}\,,
\end{equation}
where the power of the bracket is understood as a matrix power, for example for $n=2$ it is $(\partial^\mu\partial_{\alpha_1}\phi \partial^{\alpha_1} \partial_{\alpha_2}\phi
\partial^{\alpha_2} \partial_{\alpha_3}\phi
\partial^{\alpha_3} \partial^{\nu}\phi)$.
In the equivalent dual form (now with the kinetic terms dropped)
\begin{align}
\mathcal{L}_{spec}^{\prime \text{int}} &=\sum\limits_{n=2}^{4} 
 \frac{1}{(n+1)!(4-n)!}\left(\frac{2}{\alpha }\right) ^{n-1}\varphi (x)\,\varepsilon ^{\mu _{1}\mu _{2}\mu _{3}\mu _{4}}\varepsilon ^{\nu _{1}\nu
_{2}\nu _{3}\nu _{4}}\prod\limits_{i=1}^{n}\partial _{\mu _{i}}\partial _{\nu _{i}}\varphi (x)\prod\limits_{j=n+1}^{4}\eta _{\mu _{j}\nu _{j}}\notag\\
&+\frac{1}{4\alpha }\varepsilon ^{\alpha
\beta \mu _{1}\mu _{2}}\varepsilon ^{\gamma \delta \nu _{1}\nu
_{2}}F_{\alpha \beta }F_{\gamma \delta }\left( \partial
_{\mu _{1}}\partial _{\nu _{1}}\varphi \eta _{\mu _{2}\nu _{2}}+%
\frac{1}{\alpha }\partial _{\mu _{1}}\partial _{\nu _{1}}\varphi \partial _{\mu _{2}}\partial _{\nu _{2}}\varphi \right)\,.
\end{align}
It is expressed using the finite number of terms, though, the price to pay is the existence of the three-point and five-point vertices in the scalar sector.

\section{Conclusions}

In this paper, we have analyzed a particular class of effective field theories using the direct methods based on the soft amplitude bootstrap, and the Lagrangian methods based on geometry and symmetry considerations. We focused on effective field theories with spontaneous symmetry breaking which describe the dynamics of Goldstone bosons living in the cosets furnished by the broken symmetry pattern. Apart from previous works, we included the interactions with an odd number of fields which naturally lead to the non-vanishing soft theorems. 

The non-vanishing soft theorems generalize the standard Adler zero. The naive connection between the Goldstone bosons and the vanishing soft limit can be violated due to the presence of cubic vertices in Lagrangian and/or linear terms present in the non-linear symmetry transformation of the Goldstone fields. In fact, these two conditions are related. We can remove the three-point vertices by means of the field redefinition, which seems natural as there are typically no three-point amplitudes for the EFT power-counting, but we might change the symmetry transformation. 

In this paper, we have presented a generalization of the whole class of soft theorems with the non-zero right-hand side, particular cases of which have been discussed in the literature in context of photons, gluons and gravitons \cite{Elvang:2016qvq}, nonlinear sigma models \cite{Kampf:2019mcd} and multi-flavor Galileons \cite{Kampf:2020tne}.
They connect the $n$-pt amplitude with $(n-1)$-pt amplitudes multiplied by  soft factors whose form is dictated by the little-group scaling and power counting.
We have demonstrated the usefulness of the non-trivial soft theorems on a  problem of the non-existence of the vector analog of Galileon theories. 

We have first presented the amplitude-based proof of a no-go theorem for existence of a unique exceptional pure vector theory with a Galileon power counting and with enhanced vanishing soft limits.  We have also shown that the natural extension using an additional scalar degree of freedom cannot solve this problem. The solution is to require the generalized soft theorems instead of the enhanced Adler zero condition.

We have then proceeded with a formal proof of the existence of scalar-vector Galileon-like theory using a combination of the soft amplitude bootstrap  and   symmetry based construction of the Lagrangian. We have obtained a  theory with the Galileon power counting which couples  massless scalar and vector degrees of freedom and which is fixed by means of generalized soft theorems. This {\emph{general scalar-vector Galileon}}  has six parameters in total. It represents a whole class of theories known from the literature \cite{Heisenberg:2017mzp,Bonifacio:2019hrj}. These theories usually arise in connection with the decoupling limit of the massive spin-2 particle or generalized Proca fields. The status of variants was unclear, and our work thus represents a unified way of their definition.

In the scalar case, for a particular values of the parameters of the general Galileon theory, there is a unique theory, the Special Galileon, with enhanced soft limit caused by an underlying hidden symmetry. We showed that the same phenomenon also happens for scalar-vector Galileons. In this case, the requirement of the enhanced $O(p^3)$ soft scalar behavior together with generalized soft vector theorem fixes uniquely a {\emph{Special scalar-vector Galileon theory}}.
This theory can be understood as a particular case of the coupling of the Special Galileon to spin-1 massless particles and its soft limit behavior can be derived from the extended  Special Galileon symmetry and a shift symmetry of the vector field.

The above results give a hope that the general soft theorems might open the space for systematic exploration of the landscape of effective field theories with particles with various spins.

\acknowledgments

We would like to thank James Bonifacio, Lavinia Heisenberg and Congkao Wen for valuable discussions and suggestions. This work is supported in part by the Czech Government project GA \v{C}R 21-26574S, Charles University Grant Agency, project No. 1108120, Ministry of Education grant LTAUSA17069, by the U.S. Department of Energy grant No. DE-SC0009999 and the funds of University of California.

\appendix

\section{Proof of the soft theorems\label{soft_theorems}}

In this section we will give the proof of the soft theorems for the theory described by the general Lagrangian (\ref{general Galileon Lagrangian}). It is useful to rewrite the Lagrangian (\ref{general Galileon Lagrangian}) in terms of the symmetric spinors
 $\phi_{AB}$ and $\overline{\phi}_{\overset{.}{A}\overset{.}{B}}$ defined as
\begin{equation}
F_{A\overset{.}{A}B\overset{.}{B}}=F_{\mu \nu }\overline{\sigma }_{A\overset{%
.}{A}}^{\mu }\overline{\sigma }_{B\overset{.}{B}}^{\nu }=\phi
_{AB}\varepsilon _{\overset{.}{A}\overset{.}{B}}+\overline{\phi }_{\overset{.%
}{A}\overset{.}{B}}\varepsilon _{AB},
\end{equation}
where in our convention $\overline{\sigma }^{\mu }=\left( \mathbf{1},-\sigma
^{i}\right) $,
  $\sigma^{\mu}=\left(\mathbf{1},\sigma
^{i}\right)$ and 
 $ \varepsilon _{12}=\varepsilon _{\overset{.}{1}\overset{.}{2}%
}=1$, 
and to construct the Feynman rules directly in terms if the helicity spinors. Then the propagators of the fields $\phi _{AB}$ and $\overline{\phi }_{%
\overset{.}{A}\overset{.}{B}}$ which correspond to the internal lines of the Feynman graphs are
\begin{eqnarray}
\langle 0|T\phi _{AB}\left( x\right) \overline{\phi }_{\overset{.}{C}\overset%
{.}{D}}\left( y\right) |0\rangle  &=&\int \frac{\mathrm{d}^{4}p}{\left( 2\pi
\right) ^{4}}\mathrm{e}^{-\mathrm{i}p\cdot \left( x-y\right) }\frac{\mathrm{i%
}}{p^{2}+\mathrm{i}0}\left[ p_{A\overset{.}{C}}p_{B\overset{.}{D}}+p_{A%
\overset{.}{D}}p_{B\overset{.}{C}}\right]  \\
\langle 0|T\phi _{AB}\left( x\right) \phi _{CD}\left( y\right) |0\rangle  &=&-%
\mathrm{i}\int \frac{\mathrm{d}^{4}p}{\left( 2\pi \right) ^{4}}\mathrm{e}^{-%
\mathrm{i}p\cdot \left( x-y\right) }\left[ \varepsilon _{AC}\varepsilon
_{BD}+\varepsilon _{AD}\varepsilon _{BC}\right]  \\
\langle 0|T\phi _{\overset{.}{A}\overset{.}{B}}\left( x\right) \phi _{\overset{.}{C}%
\overset{.}{D}}\left( y\right) |0\rangle  &=&-\mathrm{i}\int \frac{\mathrm{d}%
^{4}p}{\left( 2\pi \right) ^{4}}\mathrm{e}^{-\mathrm{i}p\cdot \left(
x-y\right) }\left[ \varepsilon _{\overset{.}{A}\overset{.}{C}}\varepsilon _{%
\overset{.}{B}\overset{.}{D}}+\varepsilon _{\overset{.}{A}\overset{.}{D}%
}\varepsilon _{\overset{.}{B}\overset{.}{C}}\right] 
\end{eqnarray}
and we attached the polarization bispinor $\sqrt{2}|p]^{\overset{.}{%
A}}|p]^{\overset{.}{B}}$ to each helicity plus external line with momentum $p$ and similarly $\sqrt{2}|p\rangle _{A}|p\rangle _{B}$ for the helicity
minus external line\footnote{We take all particles as outgoing. The alternative convention which associates the polarization bispinors  ${\rm{i}}\sqrt{2}|p]^{\overset{.}{%
A}}|p]^{\overset{.}{B}}$ and ${\rm{i}}\sqrt{2}|p\rangle _{A}|p\rangle _{B}$ to the external lines is also possible. In this case, the sign of the parameter $a$ of the soft-vector theorem is changed.}. 
The interaction vertices can be then read off form the Lagrangian
\begin{eqnarray}
\mathcal{L} &=&\mathcal{L}_{gal}-\frac{a_{3}}{8\Lambda ^{3}}\left[ \phi
_{AB}\phi ^{AB}\square \varphi +\overline{\phi }_{\overset{.}{A}\overset{.}{B%
}}\overline{\phi }^{\overset{.}{A}\overset{.}{B}}\square \varphi +2\phi _{AB}%
\overline{\phi }_{\overset{.}{A}\overset{.}{B}}\partial ^{A\overset{.}{A}%
}\partial ^{B\overset{.}{B}}\varphi \right]  \notag\\
&&-\frac{a_{4}}{16\Lambda ^{6}}\left[ \phi _{MA}\phi _{NB}\partial _{\overset%
{.}{A}}^{~~~M}\partial _{\overset{.}{B}}^{~~~N}\varphi -2\phi _{MA}\overline{%
\phi }_{\overset{.}{N}\overset{.}{B}}\partial _{\overset{.}{A}%
}^{~~~M}\partial _{~~~B}^{\overset{.}{N}}\varphi \right. \notag \\
&&\phantom{+\frac{a_{4}}{16\Lambda ^{6}}}\left. +\overline{\phi }_{\overset{.}{M}\overset{.}{A}}\overline{\phi }_{%
\overset{.}{N}\overset{.}{B}}\partial _{~~~A}^{\overset{.}{M}}\partial
_{~~~B}^{\overset{.}{N}}\varphi \right] \partial ^{A\overset{.}{A}}\partial
^{B\overset{.}{B}}\varphi \,.
\end{eqnarray}%

\subsection{Soft scalar theorem}
\label{ap:secsst}

\begin{figure}[tb]
  \centering
    \includegraphics[scale=0.65]{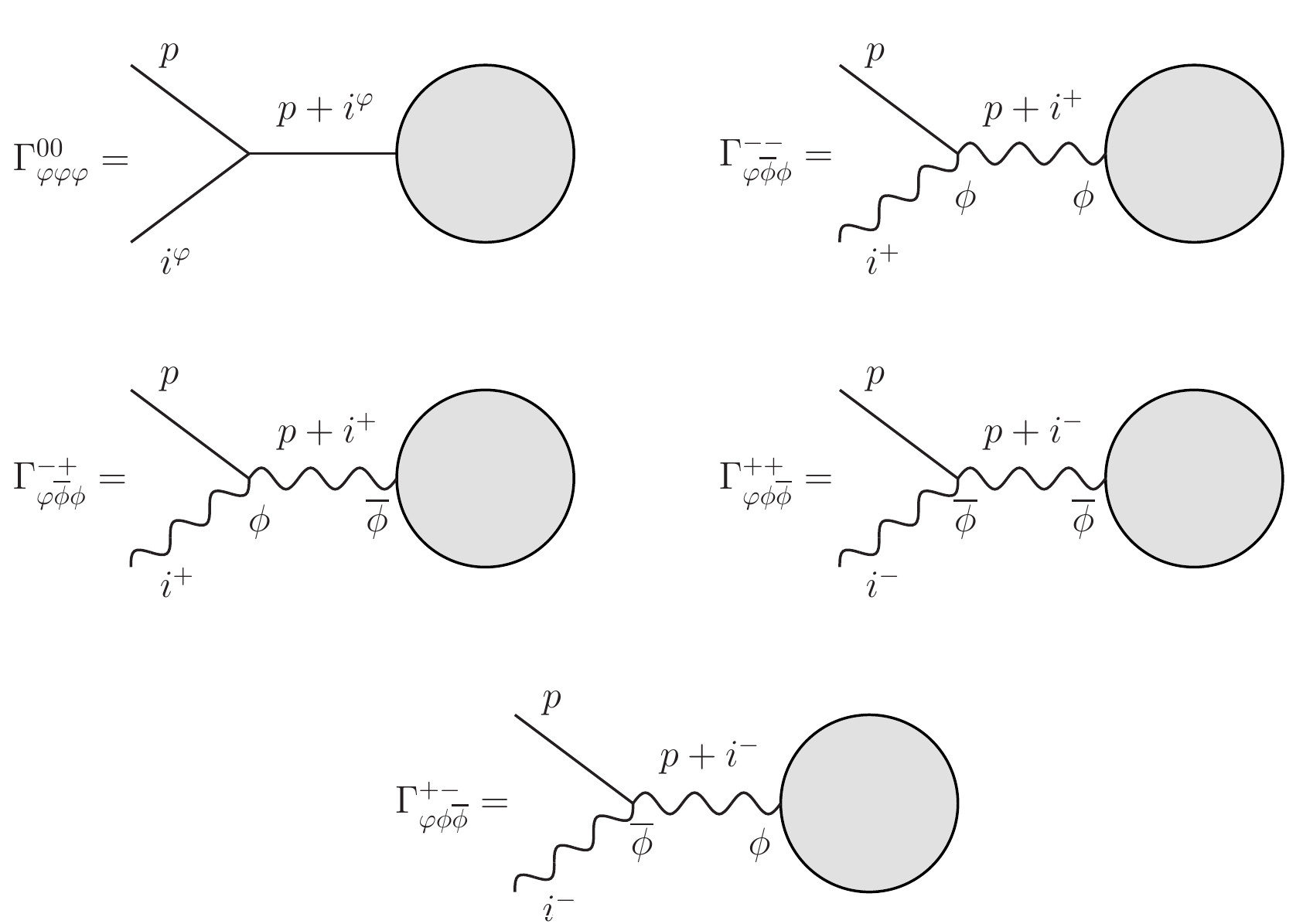}
    \caption{The graphs with a nontrivial single scalar soft limit}
    \label{fig:soft_phi}
\end{figure}

The leading order contributions in the single scalar soft limit $p\rightarrow 0$ of the amplitude $A_{n+1}\left( p^{\varphi
},I^{+},I^{-},I^{\varphi }\right) $ comes from the graphs with the soft
scalar attached to  external lines of some semi-on-shell $n$-point
amplitude. Therefore we need the semi-on-shell 3pt vertices
\begin{eqnarray}
&&V_{3\varphi }\left( p^{\varphi },i^{\varphi },-\left( p+i\right) ^{\varphi
}\right)  =12\frac{d_{3}}{\Lambda^3}\left( p\cdot i\right) ^{2}  \notag \\
&&V_{\varphi \phi \phi }\left( p^{\varphi },i^{-},-\left( p+i\right)
^{-}\right) ^{AB} =0 \notag\\
&&V_{\varphi \overline{\phi }\overline{\phi }}\left( p^{\varphi
},i^{+},-\left( p+i\right) ^{+}\right) _{\overset{.}{A}\overset{.}{B}} =0 
\notag \\
&&V_{\varphi \overline{\phi }\phi }\left( p^{\varphi },i^{+},-\left(
p+i\right) ^{-}\right) ^{AB} =\frac{a_{3}}{4\Lambda ^{3}}\sqrt{2}%
[i|p^{A}[i|p^{B}  \notag \\
&&V_{\varphi \phi \overline{\phi }}\left( p^{\varphi },i^{-},-\left(
p+i\right) ^{+}\right) _{\overset{.}{A}\overset{.}{B}} =\frac{a_{3}}{%
4\Lambda ^{3}}\sqrt{2}\langle i|p_{\overset{.}{A}}\langle i|p_{\overset{.}{B}}\,.
\end{eqnarray}
Let us denote also
\begin{equation}
    \Delta(P)_{AB\overset{.}{A}\overset{.}{B}}=\frac{P_{A\overset{.}{A}}P_{B\overset{.}{B}}+P_{A\overset{.}{B}}P_{B\overset{.}{A}}}{P^2}.
\end{equation}
The corresponding nontrivial contributions are then (see Fig.~\ref{fig:soft_phi})
\begin{eqnarray}
\Gamma_{\varphi\varphi\varphi}^{00}&=&-V_{3\varphi }\left( p^{\varphi },i^{\varphi },-\left( p+i\right)
^{\varphi }\right) \frac{1}{\left( p+i\right) ^{2}}\widetilde{A}_{n}\left(
I^{+},I^{-},I^{\varphi }\right) _{i^{\varphi }\rightarrow \left( p+i\right)
^{\varphi }} \notag\\
&=&-6\frac{d_{3}}{\Lambda^3}\left( p\cdot i\right) \widetilde{A}_{n}\left(
I^{+},I^{-},I^{\varphi }\right) _{i^{\varphi }\rightarrow \left( p+i\right)
^{\varphi }}=-6\frac{d_{3}}{\Lambda^3}\left( p\cdot i\right) A_{n}\left(
I^{+},I^{-},I^{\varphi }\right) +O\left( p^{2}\right)  \notag\\
\\
\Gamma_{\varphi \overline{\phi }\phi }^{--}&=&V_{\varphi \overline{\phi }\phi }\left( p^{\varphi },i^{+},-\left(
p+i\right) ^{-}\right) ^{AB}\left[ \varepsilon _{AC}\varepsilon
_{BD}+\varepsilon _{AD}\varepsilon _{BC}\right] \widetilde{A}_{n}\left(
I^{+},I^{-},I^{\varphi }\right) _{i^{+}\rightarrow \left( p+i\right)
^{-}}^{CD}\notag \\
&=&\frac{a_{3}}{2\Lambda ^{3}}\sqrt{2}[i|p_{A}[i|p_{B}\widetilde{A}%
_{n}\left( I^{+},I^{-},I^{\varphi }\right) _{i^{+}\rightarrow \left(
p+i\right) ^{-}}^{AB}=O\left( p^{2}\right)  \\\notag\\
\Gamma_{\varphi \overline{\phi }\phi }^{-+}&=&-V_{\varphi \overline{\phi }\phi }\left( p^{\varphi },i^{+},-\left(
p+i\right) ^{-}\right) ^{AB}
\Delta(p+i)_{AB\overset{.}{C}\overset{.}{D}}
\widetilde{A}_{n}\left( I^{+},I^{-},I^{\varphi
}\right) _{i^{+}\rightarrow \left( p+i\right) ^{+}}^{\overset{.}{C}\overset{.%
}{D}}\notag \\
&=&-\frac{a_{3}}{\Lambda ^{3}}\left( p\cdot i\right) \sqrt{2}[i|_{\overset{.}%
{C}}[i|_{\overset{.}{D}}\widetilde{A}_{n}\left( I^{+},I^{-},I^{\varphi
}\right) _{i^{+}\rightarrow \left( p^{\varphi }+i\right) ^{+}}^{\overset{.}{C%
}\overset{.}{D}}\notag\\
&=&-\frac{a_{3}}{\Lambda ^{3}}\left( p\cdot i\right)
A_{n}\left( I^{+},I^{-},I^{\varphi }\right) +O\left( p^{2}\right)  \\\notag\\
\Gamma_{\varphi \phi \overline{\phi }}^{++}&=&V_{\varphi \phi \overline{\phi }}\left( p^{\varphi },i^{-},-\left(
p+i\right) ^{+}\right) ^{\overset{.}{A}\overset{.}{B}}\left[ \varepsilon _{%
\overset{.}{A}\overset{.}{C}}\varepsilon _{\overset{.}{B}\overset{.}{D}%
}+\varepsilon _{\overset{.}{A}\overset{.}{D}}\varepsilon _{\overset{.}{B}%
\overset{.}{C}}\right] \widetilde{A}_{n}\left( I^{+},I^{-},I^{\varphi
}\right) _{i^{-}\rightarrow \left( p+i\right) ^{+}}^{\overset{.}{C}\overset{.%
}{D}}\notag \\
&=&\frac{a_{3}}{2\Lambda ^{3}}\sqrt{2}\langle i|p_{\overset{.}{A}}\langle
i|p_{\overset{.}{B}}\widetilde{A}_{n}\left( I^{+},I^{-},I^{\varphi }\right)
_{i^{-}\rightarrow \left( p+i\right) ^{+}}^{\overset{.}{C}\overset{.}{D}%
}=O\left( p^{2}\right)  \\\notag\\
\Gamma_{\varphi \phi \overline{\phi }}^{+-}&=&-V_{\varphi \phi \overline{\phi }}\left( p^{\varphi },i^{-},-\left(
p+i\right) ^{+}\right) ^{\overset{.}{A}\overset{.}{B}}
\Delta(p+i)_{\overset{.}{A}\overset{.}{B}CD}
\widetilde{A}_{n}\left(
I^{+},I^{-},I^{\varphi }\right) _{i^{-}\rightarrow \left( p+i\right)
^{-}}^{CD}\notag \\
&=&-\frac{a_{3}}{\Lambda ^{3}}\left( p\cdot i\right) \sqrt{2}\langle
i|_{C}\langle i|_{D}\widetilde{A}_{n}\left( I^{+},I^{-},I^{\varphi }\right)
_{i^{-}\rightarrow \left( p+i\right) ^{-}}^{CD}
\notag\\
&=&-\frac{a_{3}}{\Lambda ^{3}}%
\left( p\cdot i\right) A_{n}\left( I^{+},I^{-},I^{\varphi }\right)
+O\left( p^{2}\right) \,.
\end{eqnarray}
On the left hand side, we have indicated the type of the 3pt vertex involved and the superscript refers to the type of the propagator corresponding to the internal line connecting the 3pt vertex with the remnant of the graph.
On the right hand side, the tilde means that the corresponding amplitude is semi-on-shell and the subscript indicates the type of the off-shell line.

All other graphs, i.e. those where the soft scalar is attached to the 4pt
vertex or to doubly-off-shell 3pt one, yield contributions of the order $O(p^{2})$ since these vertices scale as $O(p^{2})$ and
there is no singular propagator which could cancel this behavior in the soft
limit. Summarily we get the single scalar soft theorem in the form
\begin{equation}
A_{n+1}\left( p^{\varphi },I^{+},I^{-},I^{\varphi }\right) \overset{%
p\rightarrow 0}{=}\left[ -6\frac{d_{3}}{\Lambda^3}\sum\limits_{i\in I^{\varphi }}\left(
p\cdot i\right) -\frac{a_{3}}{\Lambda ^{3}}\sum\limits_{i\in I^{+}\cup
I^{-}}\left( p\cdot i\right) \right] A_{n}\left( I^{+},I^{-},I^{\varphi
}\right) +O\left( p^{2}\right). 
\end{equation}

\subsection{Soft vector theorem}

\begin{figure}[tb]
  \centering
    \includegraphics[scale=0.65]{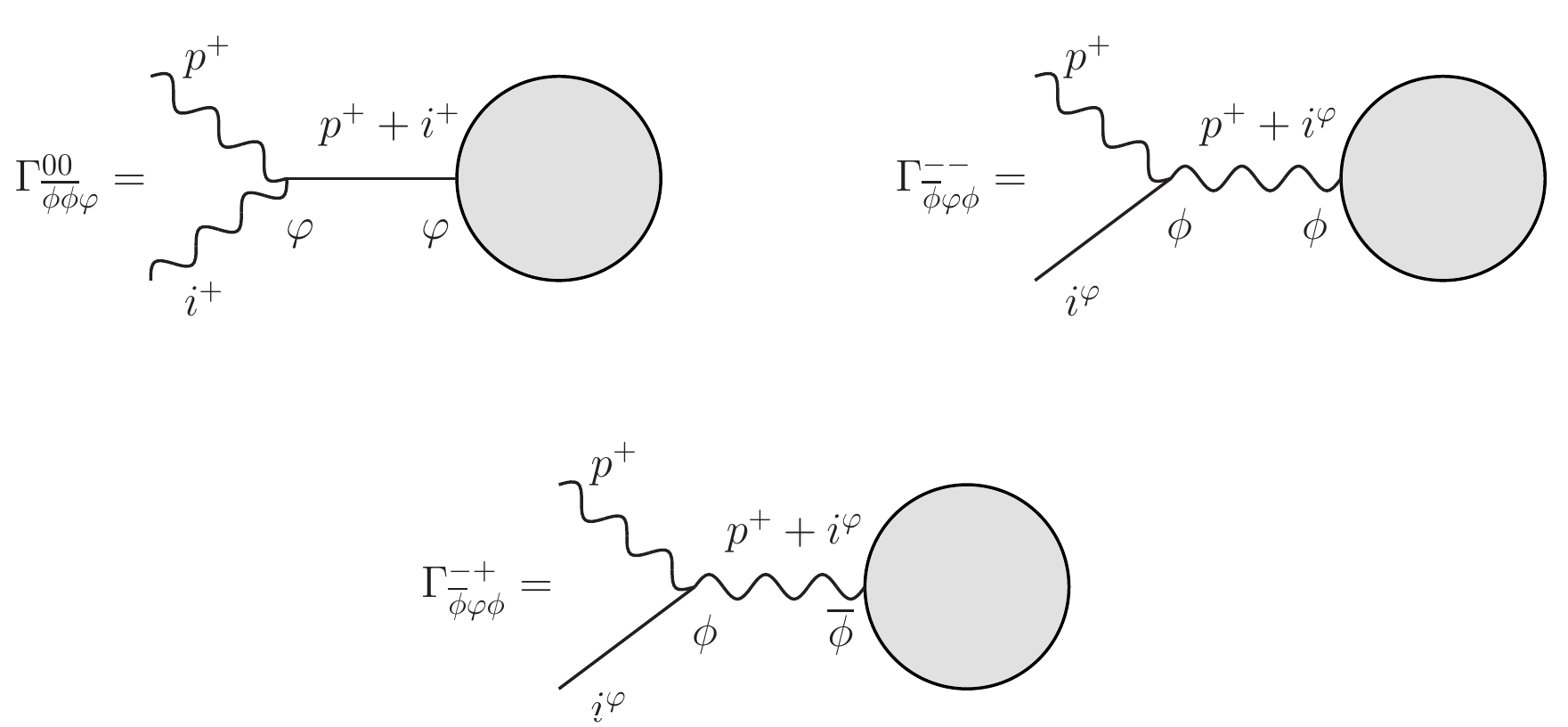}
    \caption{The contribution of graphs with a nontrivial single vector soft limit}
    \label{fig:soft_gamma}
\end{figure}

\begin{figure}[tb]
  \centering
    \includegraphics[scale=0.65]{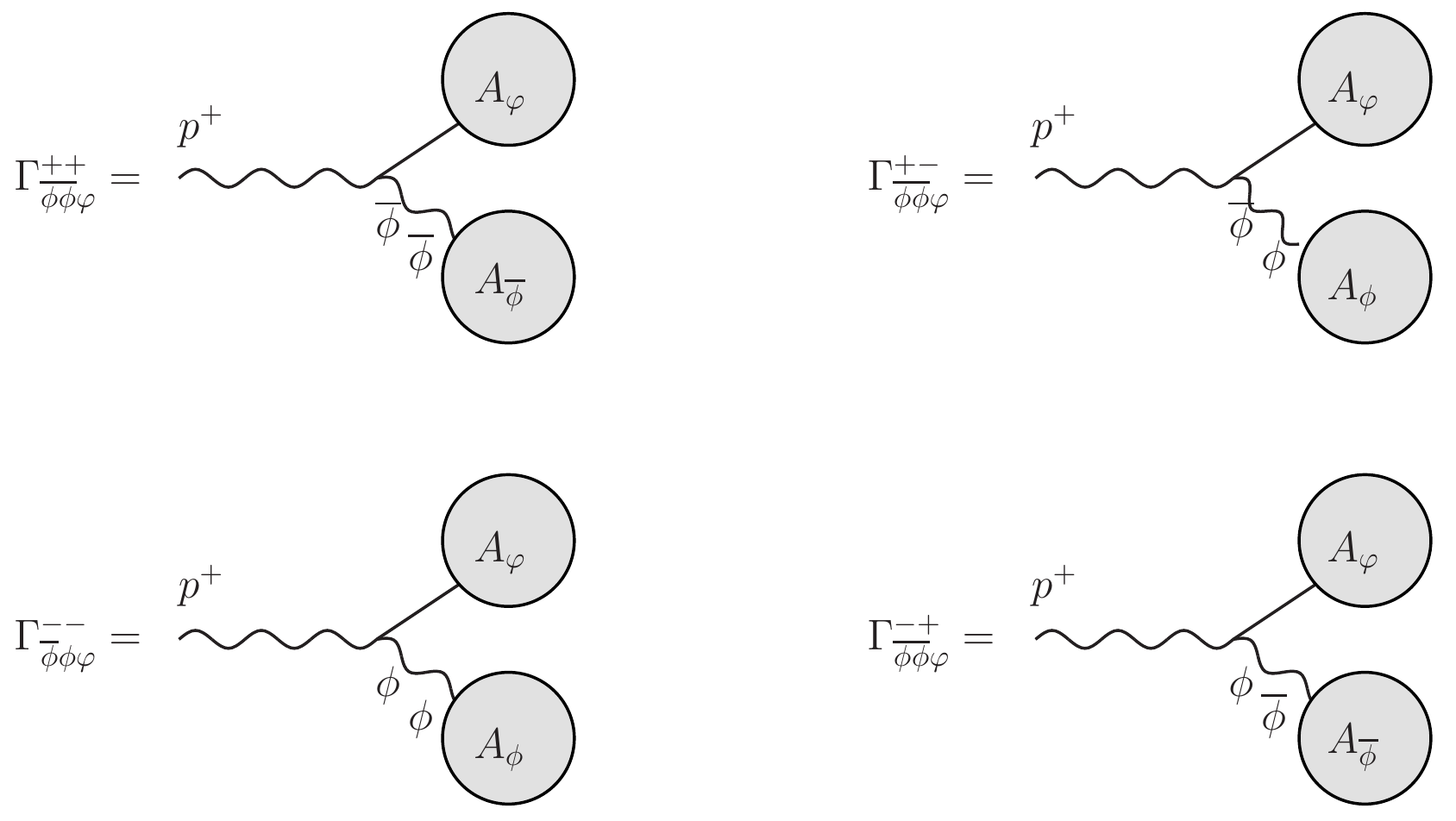}
    \caption{The contribution of graphs with a double off shell 3pt vertices with nontrivial single vector soft limit.}
    \label{fig:soft_gamma_3pt}
\end{figure}

Also in this case, the only relevant graphs are the ones with the soft vector particle attached to the external lines. For the helicity plus soft vector we need the following semi-on-shell vertices
\begin{eqnarray}
&&V_{\overline{\phi }\overline{\phi }\varphi }\left( p^{+},i^{+},-\left(
p+i\right) ^{\varphi }\right)  =\frac{a_{3}}{2\Lambda ^{3}}[p|i]^{2}\left(
p+i\right) ^{2}   \\
&&V_{\overline{\phi }\varphi \overline{\phi }}\left( p^{+},i^{\varphi
},-\left( p+i\right) ^{+}\right)  =0 \\
&&V_{\overline{\phi }\phi \varphi }\left( p^{+},i^{-},-\left( p+i\right)
^{\varphi }\right)  =\frac{a_{3}}{4\Lambda ^{3}}[p|\left( p+i\right)
|i\rangle ^{2}=0 \\
&&V_{\overline{\phi }\varphi \phi }\left( p^{+},i^{\varphi },-\left(
p+i\right) ^{-}\right) ^{AB} =\frac{a_{3}}{4\Lambda ^{3}}\sqrt{2}%
[p|i^{A}[p|i^{B}\,.
\end{eqnarray}
The corresponding nontrivial contributions are then (see Fig.~\ref{fig:soft_gamma})
\begin{eqnarray}
\Gamma _{\overline{\phi }\overline{\phi }\varphi }^{00} &=&-V_{\overline{%
\phi }\overline{\phi }\varphi }\left( p^{+},i^{+},-\left( p+i\right)
^{\varphi }\right) \frac{1}{\left( p+i\right) ^{2}}\widetilde{A}_{n}\left(
I^{+},I^{-},I^{\varphi }\right) _{i^{+}\rightarrow \left( i+p\right)
^{\varphi }}\notag \\
&=&-\frac{a_{3}}{2\Lambda ^{3}}[p|i]^{2}\widetilde{A}_{n}\left(
I^{+},I^{-},I^{\varphi }\right) _{i^{+}\rightarrow \left( i+p\right)
^{\varphi }} \notag\\
&=&-\frac{a_{3}}{2\Lambda ^{3}}[p|i]^{2}A_{n}\left( I^{+},I^{-},I^{\varphi
}\right) _{i^{+}\rightarrow i^{\varphi }}+O\left( p\right)  \\
\Gamma _{\overline{\phi }\varphi \phi }^{--} &=&V_{\overline{\phi }\varphi
\phi }\left( p^{+},i^{\varphi },-\left( p+i\right) ^{-}\right) ^{AB}\left[
\varepsilon _{AC}\varepsilon _{BD}+\varepsilon _{AD}\varepsilon _{BC}\right] 
\widetilde{A}_{n}\left( I^{+},I^{-},I^{\varphi }\right) _{i^{\varphi
}\rightarrow \left( p+i\right) ^{-}}^{CD}\notag \\
&=&\frac{a_{3}}{2\Lambda ^{3}}\sqrt{2}[p|i_{A}[p|i_{B}\widetilde{A}%
_{n}\left( I^{+},I^{-},I^{\varphi }\right) _{i^{\varphi }\rightarrow \left(
p+i\right) ^{-}}^{AB} \notag\\
&=&\frac{a_{3}}{2\Lambda ^{3}}[p|i]^{2}A_{n}\widetilde{A}_{n}\left(
I^{+},I^{-},I^{\varphi }\right) _{i^{\varphi }\rightarrow
i^{-}}^{AB}+O\left( p\right)  \\
\Gamma _{\overline{\phi }\varphi \phi }^{-+} &=&-V_{\overline{\phi }\varphi
\phi }\left( p^{+},i^{\varphi },-\left( p+i\right) ^{-}\right) ^{AB}\Delta
_{AB\overset{.}{A}\overset{.}{B}}\left( p+i\right) \widetilde{A}_{n}\left(
I^{+},I^{-},I^{\varphi }\right) _{i^{\varphi }\rightarrow \left( p+i\right)
+}^{\overset{.}{A}\overset{.}{B}}\notag \\
&=&\frac{a_{3}}{\Lambda ^{3}}\sqrt{2}\left( p\cdot i\right) [p|_{\overset{.}{%
A}}[p|_{\overset{.}{B}}\widetilde{A}_{n}\left( I^{+},I^{-},I^{\varphi
}\right) _{i^{\varphi }\rightarrow \left( p+i\right) +}^{\overset{.}{A}%
\overset{.}{B}}=O\left( p\right) \,.
\end{eqnarray}
In contrast with the scalar soft theorem, in this case the remaining
graphs can have nonzero contribution for $p\rightarrow 0$, but the
shift symmetry forces them to cancel each other. We will illustrate the
mechanism of this cancellation on the graphs where the helicity plus soft
vector particle is attached to doubly-off-shell 3pt vertices listed below:
\begin{eqnarray*}
V_{\overline{\phi }\overline{\phi }\varphi }\left( p^{+},i^{+},-\left(
p+i\right) ^{\varphi }\right) _{\overset{.}{A}\overset{.}{B}} &=&\frac{a_{3}%
}{4\Lambda ^{3}}\sqrt{2}[p|_{\overset{.}{A}}[p|_{\overset{.}{B}}\left(
p+i\right) ^{2} \\
V_{\overline{\phi }\phi \varphi }\left( p^{+},i^{-},-\left( p+i\right)
^{\varphi }\right) ^{AB} &=&\frac{a_{3}}{4\Lambda ^{3}}\sqrt{2}%
[p|i^{A}[p|i^{B}\,.
\end{eqnarray*}
We get the following contributions (see Fig.~\ref{fig:soft_gamma_3pt})
\begin{eqnarray}
\Gamma _{\overline{\phi }\overline{\phi }\varphi }^{++} &=&-V_{\overline{%
\phi }\overline{\phi }\varphi }\left( p^{+},i^{+},-\left( p+i\right)
^{\varphi }\right) ^{\overset{.}{A}\overset{.}{B}}\left[ \varepsilon _{%
\overset{.}{A}\overset{.}{C}}\varepsilon _{\overset{.}{B}\overset{.}{D}%
}+\varepsilon _{\overset{.}{A}\overset{.}{D}}\varepsilon _{\overset{.}{B}%
\overset{.}{C}}\right] \widetilde{A}^{\overset{.}{C}\overset{.}{D}}\left(
i\right) \frac{1}{\left( p+i\right) ^{2}}\widetilde{A}_{\varphi }\left(
\left( p+i\right) ^{\varphi }\right) \notag \\
&=&-\frac{a_{3}}{2\Lambda ^{3}}\sqrt{2}[p|_{\overset{.}{A}}[p|_{\overset{.}{B%
}}\widetilde{A}^{\overset{.}{A}\overset{.}{B}}\left( i^{+}\right) \widetilde{%
A}_{\varphi }\left( p+i\right)  \\
\Gamma _{\overline{\phi }\overline{\phi }\varphi }^{+-} &=&V_{\overline{\phi 
}\overline{\phi }\varphi }\left( p^{+},i^{+},-\left( p+i\right) ^{\varphi
}\right) ^{\overset{.}{A}\overset{.}{B}}\Delta \left( i\right) _{\overset{.}{%
A}\overset{.}{B}CD}\widetilde{A}^{CD}\left( i^{-}\right) \frac{1}{\left(
p+i\right) ^{2}}\widetilde{A}_{\varphi }\left( \left( p+i\right) ^{\varphi
}\right) \notag \\
&=&\frac{a_{3}}{2\Lambda ^{3}}\sqrt{2}[p|i_{A}[p|i_{B}\widetilde{A}%
^{AB}\left( i^{-}\right) \frac{1}{ i^{2}}\widetilde{A}%
_{\varphi }\left( \left( p+i\right) ^{\varphi }\right)  \\
\Gamma _{\overline{\phi }\phi \varphi }^{--} &=&-V_{\overline{\phi }\phi
\varphi }\left( p^{+},i^{-},-\left( p+i\right) ^{\varphi }\right) ^{AB}\left[
\varepsilon _{AC}\varepsilon _{BD}+\varepsilon _{AD}\varepsilon _{BC}\right] 
\widetilde{A}^{CD}\left( i\right) \frac{1}{\left( p+i\right) ^{2}}\widetilde{%
A}_{\varphi }\left( \left( p+i\right) ^{\varphi }\right) \notag \\
&=&-\frac{a_{3}}{2\Lambda ^{3}}\sqrt{2}[p|i_{A}[p|i_{B}\widetilde{A}%
^{AB}\left( i\right) \frac{1}{\left( p+i\right) ^{2}}\widetilde{A}_{\varphi
}\left( \left( p+i\right) ^{\varphi }\right)  \\
\Gamma _{\overline{\phi }\phi \varphi }^{-+} &=&V_{\overline{\phi }\phi
\varphi }\left( p^{+},i^{-},-\left( p+i\right) ^{\varphi }\right)
^{AB}\Delta \left( i\right) _{AB\overset{.}{C}\overset{.}{D}}\widetilde{A}^{%
\overset{.}{C}\overset{.}{D}}\left( i\right) \frac{1}{\left( p+i\right) ^{2}}%
\widetilde{A}_{\varphi }\left( \left( p+i\right) ^{\varphi }\right) \notag \\
&=&\frac{a_{3}}{4\Lambda ^{3}}\sqrt{2}[p|_{\overset{.}{A}}[p|_{\overset{.}{B}%
}\widetilde{A}^{\overset{.}{A}\overset{.}{B}}\left( i^{+}\right) \widetilde{A%
}_{\varphi }\left( p+i\right) \frac{i^2}{(p+i)^2}
\end{eqnarray}
and thus the contributions, which differ only by the replacement of the contact helicity violating propagator with the pole helicity conserving propagator but with the same off-shell remnants $\widetilde{A}_{\varphi}$ and $\widetilde{A}^{AB}$ or $\widetilde{A}^{\overset{.}{A} \overset{.}{B}}$, cancel each other in the limit $p\to 0$. Namely,
\begin{equation*}
\lim_{p\to 0}\left(\Gamma _{\overline{\phi }\overline{\phi }\varphi }^{++}+\Gamma _{\overline{%
\phi }\phi \varphi }^{-+}\right)=0,~~~~\lim_{p\to 0}\left(\Gamma _{\overline{\phi }\overline{\phi }%
\varphi }^{+-}+\Gamma _{\overline{\phi }\phi \varphi }^{--}\right)=0.
\end{equation*}
A similar mechanism ensures cancellation of possible nonzero contributions of those graphs where the soft vector leg is attached to the (semi)off-shell 4pt vertex.
Summing up therefore only the non-vanishing contributions in the soft limit, we get the soft theorem in the form
\begin{eqnarray}
A_{n+1}\left( p^{+},I^{+},I^{-},I^{\varphi }\right) &\overset{%
p\rightarrow 0}{=}&-%
\frac{a_3}{2\Lambda ^{3}}\sum\limits_{i\in I^{+}}\left[ p,i\right]
^{2}A_{n}\left( I^{+},I^{-},I^{\varphi }\right) _{i^{+}\rightarrow
i^{\varphi }}\notag\\
&&+\frac{a_3}{2\Lambda ^{3}}\sum\limits_{i\in I^{\varphi }}\left[
p,i\right] ^{2}A_{n}\left( I^{+},I^{-},I^{\varphi }\right) _{i^{\varphi
}\rightarrow i^{-}}+O(p)\,.
\label{eq:softtheorema3}
\end{eqnarray}

\section{4pt and 5pt amplitudes\label{app:amp45}}

We have presented our results in Sec.~\ref{sec:sum} using the soft-theorem parameters and the seed amplitudes. Here we will summarize the 4pt and 5pt amplitudes using the Lagrangian parameters introduced in Sec.~\ref{sec:genscalgal}. As already mentioned, we have used a certain convention which can differ from others due to the phase for a certain field type. Typically, the vector external field could have been multiplied by an extra $i$ and then one has to flip a sign for the amplitudes below with exactly two vectors.

Let us start with the parameters of the soft theorems~(\ref{eq:softx}) and~(\ref{eq:softy})
\begin{equation}
    \xi = \frac{2 a_3 - \theta}{4\Lambda^3}\,,\qquad
    \eta = \frac{-12 d_3 + 2a_3 + \theta}{2\Lambda^3}\,.
\end{equation}
The 4pt amplitudes are
\begin{align}
&\Lambda^6\,A_{04}(1^\varphi2^\varphi3^\varphi4^\varphi) = -3 (9 d_3^2 -2 d_4) \, s_{12}s_{13}s_{23}\notag\\
&\Lambda^6\,A_{22}(1^+2^-3^\varphi4^\varphi) = -\frac{1}{16} ( (2 a_3+\theta)^2 - 8 a_4 )\, s_{34}[1|3|2\ra[1|4|2\ra\notag\\
&\Lambda^6\,A_{22}(1^+2^+3^\varphi4^\varphi) = -\frac{1}{16} 
\bigl((2a_3- \theta)(2a_3+ \theta)  -4 a_4 \bigr)\, (s_{13} s_{23}+s_{14} s_{24})[12]^2\notag\\ &\qquad\qquad\qquad\qquad\qquad + \frac{1}{16}(2 a_3-\theta)(2 a_3+\theta-12 d_3)\, s_{12}^2 [12]^2  \notag\\
&\Lambda^6\,A_{40}(1^+2^+3^-4^-)= -\frac{1}{16} (2 a_3 - \theta)^2\, s_{12}[12]^2\la34\ra^2\notag\\
&\Lambda^6\,A_{40}(1^+2^+3^+4^+)= -\frac{1}{16} (2 a_3 - \theta)^2\, \bigl(s_{12}[12]^2[34]^2 + s_{13}[13]^2[24]^2 + s_{23} [23]^2[14]^2\bigr)\notag\\
&\Lambda^6\,A_{40}(1^+2^+3^+4^-)= 0\,.
\end{align}
The pure scalar 5pt amplitude is
\begin{align}
    &\Lambda^9\,A_{05}(1^\varphi 2^\varphi 3^\varphi 4^\varphi 5^\varphi) = -\frac{1}{16}(72 d_3^3 -24 d_3 d_4+5 d_5) s_{12}s_{13}s_{14}s_{15} + \text{perm(1,2,3,4,5)}\,.
\end{align}
It is the most important 5pt amplitude as it introduces a new and final parameter $d_5$ and all other amplitudes can be now obtained by the soft recursion. For simplicity, we will summarize the remaining 5pt amplitudes only for $d_3=d_4=a_4=0$:
\begin{align}
&\Lambda^8\,A_{23}(1^+ 2^+ 3^\varphi 4^\varphi 5^\varphi) =  -\frac{1}{64}(2a_3-\theta)(2a_3+\theta)^2 [1|45|2]^2(s_{13}+s_{23})+ \text{perm(3,4,5)}\notag\\
&\Lambda^8\,A_{23}(1^+ 2^- 3^\varphi 4^\varphi 5^\varphi) = \frac{1}{64}(2a_3 - \theta)^2(2a_3+\theta)[1|345|2\rangle^2 \notag\\&\qquad\qquad\qquad\qquad\qquad+ \frac{1}{64}(2a_3+\theta)^3 [1|4|2\rangle^2 s_{13}s_{25} + \text{perm(3,4,5)}\notag\\  
&\Lambda^8\,A_{41}(1^+ 2^+ 3^+ 4^+ 5^\varphi) =\frac{1}{64}(2a_3-\theta)^2(2a_3+\theta)[12]^2[34]^2 s_{13}s_{15}+ \text{perm(1,2,3,4)}\notag\\ 
&\Lambda^8\,A_{41}(1^+ 2^+ 3^+ 4^- 5^\varphi) =\frac{1}{64}(2a_3-\theta)^3[12]^2[3|5|4\rangle^2 s_{15}
\notag\\&\qquad\qquad\qquad\qquad\qquad
+ \frac{1}{32}(2 a_3-\theta)(4 a_3^2+\theta^2)[12]^2
[3|5|4\rangle [3|2|4\rangle s_{34}+\text{perm(1,2,3)}\notag\\
&\Lambda^8\,A_{41}(1^+ 2^+ 3^- 4^- 5^\varphi) =\frac{1}{64}(2a_3-\theta)^2(2a_3+\theta)[12]^2\langle34\rangle^2\Bigl(s_{12}(s_{35}+s_{45}) + s_{34}(s_{15}+s_{25})\notag\\& \qquad\qquad\qquad\qquad\qquad\qquad\qquad\qquad\qquad\qquad\qquad\qquad- s_{15}^2 - s_{25}^2 - s_{35}^2 -s_{45}^2\Bigr)\,.
\end{align}
\bibliographystyle{JHEP}
\bibliography{references.bib}

\end{document}